\documentclass[12pt]{article}
\usepackage{amsmath}
\usepackage{graphicx}
\usepackage{enumerate}
\usepackage{natbib}
\usepackage{amsfonts}
\usepackage{algorithm}
\usepackage{algpseudocode}
\usepackage{multirow}
\usepackage{bbm}
\usepackage{microtype}
\usepackage{supertabular}
\usepackage{hyperref}

\newcommand{\blind}{1}

\DeclareMathOperator{\CRPS}{CRPS}
\DeclareMathOperator{\Error}{Error}
\DeclareMathOperator{\Score}{Score}
\DeclareMathOperator{\logit}{logit}
\DeclareMathOperator{\normal}{normal}
\DeclareMathOperator{\E}{E}
\DeclareMathOperator{\elpd}{elpd}

\date{15 Feb 2024}

\begin{document}

\def\spacingset#1{\renewcommand{\baselinestretch}%
{#1}\small\normalsize} \spacingset{1}


\if1\blind
{
  \title{\bf Model validation for aggregate inferences in out-of-sample prediction\thanks{
    We gratefully acknowledge support from National Institutes of Health grants 5R01AG067149-02. We acknowledge the Office of Naval Research for partial support of this work. We acknowledge the support by the Research Council of Finland Flagship 
programme: Finnish Center for Artificial Intelligence, and Research Council of 
Finland project (340721). 
    We thank Daniel Simpson and Jonah Gabry for early discussions on validating models for MRP (particularly leave-one-cell-out approaches). We thank Swen Kuh for generously sharing simulation code for this work. We thank Alex Cooper for early work as a research assistant investigating scoring rules. Code for the simulations presented in this work can be found at \url{https://github.com/lauken13/MRP_validation_ALT}
    }}
  \author{Lauren Kennedy\footnote{School of Computer and Mathematical Sciences, University of Adelaide},
    Aki Vehtari\footnote{Department of Computer Science, Aalto  University},
    Andrew Gelman\footnote{Department of Statistics and Department of Political Science, Columbia University}}
  \maketitle
} \fi

\if0\blind
{
  \bigskip
  \bigskip
  \bigskip
  \begin{center}
    {\LARGE\bf  Model validation for aggregate inferences in out-of-sample prediction}
\end{center}
  \medskip
} \fi

\bigskip
\begin{abstract}
Generalization to new samples is a fundamental rationale for statistical modeling.   For this purpose, model validation is particularly important, but recent work in survey inference has suggested that simple aggregation of individual prediction scores does not give a good measure of the score for  population aggregate estimates. In this manuscript we explain why this occurs, propose two scoring metrics designed specifically for this problem, and demonstrate their use in three different ways. We show that these scoring metrics correctly order models when compared to the true score, although they do underestimate the magnitude of the score.  We demonstrate with a problem in survey research, where multilevel regression and poststratification (MRP) has been used extensively to adjust convenience and low-response surveys to make population and subpopulation estimates.
\end{abstract}

\noindent%
\vfill

\newpage

\section{Introduction}

A central challenge of statistics is to generalizing from observed data to new settings.  To do this, sample surveys increasingly rely on model-based estimates to overcome low response rates and non-probability samples. These model-based estimates use a prediction framework, but the purpose is not to predict each unobserved individual accurately. Instead the purpose is to estimate accurate population summaries from individual predictions. Although these two aims might seem similar, recent work suggests that the best model for the individual predictions might not be the best model for population summaries \citep{kuh2022using}. This leads to challenges in assessing both model adequacy and model selection \citep{gelfand1992model}. 

In this manuscript we focus on a specific model-based technique, multilevel regression and poststratification (MRP; \citealp{gelman1997, park2006}), which has been widely used in political science (e.g., \citealp{lax2009should,lax2009gay,ghitza2013,wang2015}) and increasingly deployed in social (e.g., \citealp{alsalti2023using,kolczynska2020trust}) and health sciences (e.g., \citealp{machalek2022serological,downes2018}). As with all methods of survey adjustment, the method relies on assumptions of model suitability. 

MRP has three stages (see, e.g., \citealp{lopez2022}). First, a model (originally a Bayesian multilevel model, but increasingly other regularising models, e.g., \citealp{ornstein2020,bisbee2019,gao2021}) is fit to the observed sample data conditional on discrete predictors (typically geographic and demographic) whose joint distribution in the population is known or can be estimated.  Second, a poststratification table is constructed to represent this population joint distribution, ideally using counts derived from census data or high quality official statistics. The model from the sample is then used to predict the expected value for each cell. Third, the population totals in each cell are used to weight the cell estimates to obtain an estimate for the population (or subpopulation). The three steps can be remembered as model-predict-aggregate.  

The population estimate relies on the model prediction, and so it is essential to consider how to fairly validate the model. Given the emphasis on prediction in this method, previous work has investigated the sum of leave-one-out (LOO) expected-log-predictive-density (ELPD)  \citep{kuh2022using}. This is a commonly used metric for model goodness in the Bayesian literature, and is easily implemented through the LOO package \citep{vehtari2021package}.  \cite{kuh2022using} also consider a weighted alternative to this metric (initially proposed by \citealp{lumley2015}) to account for non-random sampling. 

\cite{kuh2022using} use results from \cite{little2005does} to demonstrate that neither standard sum of LOO-ELPD nor the weighted alternative correctly distinguish between a model that correctly produces an approximately unbiased estimate of the mean and one that does not. 

\subsection{Individual scores and aggregate scores}
We argue that the underlying difficulty described in \cite{kuh2022using} is that the best model for predicting individuals may not correspond to the best model for an aggregate estimate. To understand why, consider a simple example of a sample of size 2, where these individuals have a true outcome value 0. We will say $y_{\mathrm{true}} = \{0,0\}$. The mean of $y_{true}$ is $E(y_{\mathrm{true}}) = \frac{0+0}{2} = 0$. 
Now we have two potential models to estimate $y_\mathrm{true}$. Each produces an estimate (column 2, Table~\ref{tab:pointwise_vs_mean}), from which we can calculate the squared error and the mean squared error over both individuals. We can also use these estimates to create an estimate of the mean, and from this calculate the squared error of the mean. As we can see, while model 1 is preferred by the mean squared individual error, model 2 is far preferred at estimating the mean. 

\begin{table}
    \centering
    \begin{tabular}{cccccccc}
      & \multicolumn{3}{c}{\textbf{Individual}} & &\multicolumn{2}{c}{\textbf{Mean}} \\
       & $\hat{y_i}$ & $(\hat{y_i} - y_i)^2$ & $\frac{1}{2}\sum(\hat{y_i} - y_i)^2$ && $\frac{1}{2}\sum \hat{y_i}$ & $(\frac{1}{2}\sum \hat{y_i} - \frac{1}{2}\sum y_i)^2 $ \\
      \hline
      Model 1 & \{$0,1$\} & \{$0,1$\}& 0.5 && 0.5 & 0.25 \\
      Model 2 & \{$-2,2$\} & \{$-2,2$\}& 4 && 0 & 0
    \end{tabular}
    \caption{\it Demonstration of difference between mean of individual squared error and squared error of mean in a simple example of $n=2$. True outcome $y_i=0$.}
    \label{tab:pointwise_vs_mean}
\end{table}

This simple example demonstrates what we believe to be one of the underlying challenges of validating multilevel regression and poststratification (MRP) models. An MRP based estimator is by definition an estimator that uses cellwise estimates to compose a population or subpopulation estimates. Previous investigations used sum of individual goodness to estimate the goodness of the MRP estimate, which was not effective \citep{kuh2022using}. 

\subsection{Notation and key terms}

Consider a finite population $P$ with $N$ individuals, where interest lies in a binary variable $Y$\footnote{Whilst some authors have used a continuous outcome variable \citep{liu2022}, the prevalent applications currently focus on estimating probabilities.}. Assume a set of $k$  predictors whose values are known in the population and are represented by a $N\times k$ matrix, $\mathbf{X}$.  Assume the predictors are discrete, with the $k^{th}$ variable, $X^{(k)}$, having $l^{(k)}$ levels. This means the distribution of $X$ in the population can be represented by the number of items within each of the  $J = \prod_k l^{(k)}$ possible configurations of predictors. 

The population mean of variable of interest can be written as,
\begin{equation}
\label{eqn:cell_ytrue}
    \E_{\mathrm{P}}(Y) = \frac{\sum_{i=1}^NY_i}{N} 
 = \frac{\sum_{j=1}^JN_j\mbox{Pr}(Y_{i \in j} = 1)}{N},
\end{equation}
where $\mbox{Pr}(Y_{i \in j} = 1) = \sum_{i \in j}Y_i/N_j$.

Next take a sample of size $n$. The probability of inclusion in the sample for every $j^{th}$ cell is assumed to be the same and denoted $\pi_j$.

Following traditional sampling notation, we use lower case for the sample and upper case for the population, and where a matrix is needed, it will be presented in boldface or evident from context. Our first step is to estimate the probability of the outcome in the $j^{th}$ cell in the population using the sample, denoted $\mbox{Pr}(Y_{i\in j}|y,\theta)$ as shorthand for $\mbox{Pr}(Y_{i\in j} = 1|y,\textbf{x}, \theta)$. We then use this to estimate the population mean,  $E_P(Y|y,\theta)$. We do this by modifying \eqref{eqn:cell_ytrue} into
\begin{equation}
\label{eqn:est_ytrue}
    \E_{\mathrm{P}}(Y|y,\theta) = \frac{\sum_{j=1}^JN_j\mbox{Pr}(Y_{i\in j}|y,\theta)}{\sum_{j=1}^JN_j}.
\end{equation}

Our goal is to identify the best model $M$ and estimate the parameters $\theta$, denoted $\hat{\theta}$, using the sample. We can then use this to estimate $\mbox{Pr}(Y_{ i \in j}|y,\hat{\theta})$, which can then be used in \eqref{eqn:est_ytrue}. 
\begin{equation}
\label{eqn:est_y_param}
    \E_{\mathrm{P}}(Y|y,\hat{\theta}) = \frac{\sum_{j=1}^JN_j\mbox{Pr}(Y_{i \in j}|y,\hat{\theta})}{\sum_{j=1}^JN_j}.
\end{equation}

In this paper we discuss aggregation at a range of different levels. For clarity we describe all terms with a simple example. Consider a survey of voting intention. The smallest unit is the \textit{individual}, the person who is being surveyed. These individuals can be aggregated in different ways. For example, we could aggregate all individuals with a specific set of demographic qualities (e.g., aged 18--25, female, college  educated). We call this a cell if these three demographics are the full set of $X$ variables. We could also aggregate at a particular \textit{level} (e.g., 18-25 year) of a variable (age), which we call a \textit{subpopulation} (also known as a small area). In the limit the subpopulation is the full \textit{population} (e.g., all US voters). 

\subsection{Contributions}

In this manuscript we

\begin{itemize}
    \item Propose two alternative scoring metrics for estimating the probability of a binary outcome in the population (Section 2)
    \item Demonstrate the use of this method in a leave-one-cell-out cross validation scheme, and demonstrate that unlike previous work, this method retains the correct model ordering (Section 3)
    \item Apply this method to subpopulation estimation (Section 4)
    \item Extend the leave-one-cell-out cross validation to a leave-one-cell-out reference validation score where instead of using the sample as a proxy for the cellwise population truth we use a reference model estimate (Section 5)
    \item The leave-one-cell-out approach requires all cells to be observed in the sample. We also propose an approximate leave-one-cell-out reference  approach for use when not all cells are observed in the sample (Section 5).
\end{itemize}

Together we intend for these contributions to provide a convincing argument that the proposed scoring metric is more appropriate for scoring models for MRP. 

\section {Aggregate scoring from cellwise components}
\label{sec:aggregate_scoring}

In this section we demonstrate that by decomposing a score of the MRP population mean into its cellwise components, we can produce an estimate of the population mean score using cellwise goodness. We complete this for a squared error score ($\Error_{}^2$) for the population mean and the continuous ranked probability score (CRPS). 

\subsection{Squared Error}
\label{subsection:squared_error}
First consider the square of the error for the population mean,
\begin{equation*}
    \Error_{\mathrm{MRP}} = E_P(Y|\theta, y)-E_P(Y) .
\end{equation*}
Using \eqref{eqn:cell_ytrue} and \eqref{eqn:est_y_param}, we can instead write:
\begin{equation*}
    \Error_{\mathrm{MRP}} = \ \frac{\sum_{j=1}^JN_j \mbox{Pr}(Y_{i \in j}|y,\hat{\theta})}{\sum_{j=1}^J N_j}- \frac{\sum_{j=1}^JN_j\mbox{Pr}(Y_{i \in j})}{\sum_{j=1}^J N_j}\,
\end{equation*}
which can be simplified to
\begin{equation}
\label{eqn:mrp_mse}
    \Error_{\mathrm{MRP}} = \frac{\sum_{j=1}^JN_j(\mbox{Pr}(Y_{i \in j}|y,\hat{\theta}) \,-\, \mbox{Pr}(Y_{i \in j}))}{\sum_{j=1}^J N_j}.
\end{equation}

This sum itself needs to be squared to get the squared error for the population mean

\begin{equation*}
    \Error_{\mathrm{MRP}}^2 = \Bigg(\frac{\sum_{j=1}^JN_j(\mbox{Pr}(Y_{i \in j}|y,\hat{\theta}) \,-\, \mbox{Pr}(Y_{i \in j}))}{\sum_{j=1}^J N_j}\Bigg)^2,
\end{equation*}

which is why simply summing the squared error of the individuals or cells (as is often done when scoring models; see \citealp{gelman2014understanding}) would not provide an acceptable score for the population mean. For clarity, the mean of the squared error of the cells, adjusted for sample representation would be
\begin{equation*}
    \frac{\sum_{j=1}^J N_j( \mbox{Pr}(Y_{i \in j}|y,\hat{\theta}) - \mbox{Pr}(Y_{i \in j}))^2}{\sum_{j=1}^J N_j}.
\end{equation*}

For notation simplicity we write the error instead of squared error for formulae throughout this manuscript, but figures show the squared error. 

\subsection{Continuous ranked probability scores}

The continuous ranked probability score (CRPS) is an extension of squared error that incorporates the full probability distribution. This relationship suggests that the CRPS for the population level MRP estimate be similarly decomposed. CRPS is described as 
\begin{equation*}
    \CRPS(F, y) = - \int_{-\infty}^{\infty} (F(y) - \mathbbm{1}(y \geq x))^2\,dy,
\end{equation*}
where $F$ is the cumulative distirbution function of the predictive distribution and $x$ is the true value. One benefit of this score is that it can be implemented through approximate draws of this distribution (Equation (21) of \citealp{gneiting2007strictly}). To harmonise notation, we denote $\phi$ as the posterior estimate for the population expectation and $\mathbf{\phi}'$ as a permutation of the posterior draws of $\mathbf{\phi}$, 
\begin{equation*}
    \CRPS(\phi,E_P(Y))  =  \frac{1}{2}E(\lvert\phi-\phi' \rvert) - E(\lvert \phi-E_P(Y) \rvert) . 
\end{equation*}

In our setting we focus on cases where we use posterior draws to estimate expectation. Denoting the $b^{th}$ posterior draw as $\mathbf{\phi}^{b}$, we can write this as
\begin{equation*}
    \CRPS(\phi,E_P(Y))  =  \frac{1}{2}\frac{1}{B}\sum_b^B\lvert \phi^b-\phi'^b \rvert - \frac{1}{B}\sum_b^B\lvert \phi^b-E_P(Y) \rvert.
\end{equation*}
The $\phi$ represents our MRP estimate, and $E_P(Y)$ the population mean truth, so we can rewrite  as
\begin{eqnarray*}
\nonumber
&&   \!\!\!\!\!\!\!\!  \CRPS_{\mathrm{MRP}}(Y,y,\hat{\theta})  =\\
\nonumber
&& 
\frac{1}{B}\sum_b^B\Bigg(\frac{1}{2} \ \Bigg\lvert \frac{\sum_{j=1}^J( N_j\mbox{Pr}(Y_{i\in j}|y, \hat{\theta}^b))}{\sum_{j=1}^J N_j}\, - \,\frac{\sum_{j=1}^J(N_j\mbox{Pr}(Y_{i\in j}|y, \hat{\theta}'^b))}{\sum_{j=1}^J N_j}\Bigg\rvert \,- \\
\nonumber
&& \ \ \ \ \ \ \ \
    \Bigg\lvert \frac{\sum_{j=1}^J(N_j\mbox{Pr}(Y_{i\in j}|y, \hat{\theta}^b))}{\sum_{j=1}^J N_j}\, - \,\frac{\sum_{j=1}^J(N_j\mbox{Pr}(Y_{i\in j}))}{\sum_{j=1}^J N_j} \Bigg\rvert\Bigg) .   
\end{eqnarray*}
Within each summation, this can be reordered to produce a weighted sum of cellwise components:
\begin{multline}
\label{eqn:mrp_crps}
\!\!\!\!    \CRPS_{\mathrm{MRP}}(Y,y,\hat{\theta})  =  \frac{1}{B}\sum_b^B\Bigg(\frac{1}{2} \ \Bigg\lvert \frac{\sum_{j=1}^JN_j\Big(\mbox{Pr}(Y_{i\in j}|y, \hat{\theta}^b) - \mbox{Pr}(Y_{i\in j}|y, \hat{\theta}'^b)\Big)}{\sum_{j=1}^J N_j}\Bigg\rvert  \,- \\
    \Bigg\lvert \frac{\sum_{j=1}^JN_j\Big(\mbox{Pr}(Y_{i\in j} |y, \hat{\theta}^b)- \mbox{Pr}(Y_{i \in j})\Big)}{\sum_{j=1}^J N_j} \Bigg\rvert\Bigg)  .  
\end{multline}
As with squared error, the absolute value is outside of the summation over cells, rather than inside as we would expect for a usual sum of the cellwise scores:
\begin{equation}
\label{eqn:pointwise_crps}
\begin{split}
   \CRPS(Y,y,\hat{\theta})  =  \frac{1}{N}\sum_{j=1}^JN_j\Bigg(\frac{1}{2}\frac{1}{B}\sum_b^B\Big\lvert \mbox{Pr}(Y_{i \in j}|y,\hat{\theta}^b) - \mbox{Pr}(Y_{i \in j}|y,\hat{\theta}'^b)\Big\rvert \,-\, \\
    \frac{1}{B}\sum_b^B\Big\lvert \mbox{Pr}(Y_{i \in j} |y,\hat{\theta}^b) - \mbox{Pr}(Y_{i \in j})\Big\rvert \Bigg).
\end{split}
\end{equation}

\section{Scoring rule applied with cross validation}
\label{sec:estimating_goodness}

We aim to score our model without the population truth through the observed values. We first consider a scenario where at least one individual in every cell is observed. We continue this assumption in Section~\ref{sec:sub_pop} where we consider subpopulation estimates. In Section~\ref{sec:not_all_obs} we relax this assumption with a reference validation approach. 

\subsection{Simulation Design}
For simplicity, we borrow the simulation scenario in Example 1 of \cite{kuh2022using}, but add an additional constraint that all cells in the population must be observed at least once in the sample. \cite{kuh2022using} had a less restrictive constraint that at least one observation in every level needed to be observed. To create our simulated data, first four independent $\mbox{normal}(0, 2)$ variables are sampled. The probability density of the outcome ($\mbox{Pr}(Y_{i \in j = 1})$) and the probability of being included in the sample ($\pi_{i \in j}$) were created as follows:
 \begin{align}
      &\text{Probability of outcome:  Pr}(Y_{i \in j}= 1) = \text{logit}^{-1}(0.1X_1 + 1X_2 + 0.1X_3 + 1X_4), \label{eq:yprob} \\
      &\text{Inclusion probability: }\pi_{i \in j} = \text{logit}^{-1}(0.1X_1 + 0.1X_2 + 1X_3 + 1X_4). \label{eq:inclprob}
 \end{align}
Here, $X_2$ is a precision variable (related strongly to the outcome but weakly to the probability of inclusion in the sample), and $X_4$ is a bias variable (related strongly to both). According to \cite{little2005does}, models with $X_4$ and not $X_2$ should be strongly preferred over models with $X_2$ and not $X_4$.

We create binary observations $Y_i$ for every individual in the population using $\mbox{Pr}(Y_{i \in j} = 1)$. We also discretize the $X_k$ variable into $5$ groups of equal range. 

We simulate with a population size of $20,000$, twice the size of \cite{kuh2022using}. We sample from the population by first ensuring every cell has one observation and then sampled the remaining obsevations with probability $\pi_{i \in j}$. The sample size is held constant at $1,000$. As the outcome is binary, we choose to fit a binomial likelihood of counts of $y_{i \in j} = 1$ per cell $j$. For simplicity we fit a subset of the models used by \cite{kuh2022using}, focusing on the following models:
\begin{itemize}
    \item Full model: $\mbox{Pr}(y_{i \in j} = 1|n_j) = \logit^{-1}(\beta_0 + \alpha^{(X_1)}_{j} + \alpha^{(X_2)}_{j} + \alpha^{(X_3)}_{j} + \alpha^{(X_4)}_{j}),$
    \item Precision variable model: $\mbox{Pr}(y_{i \in j} = 1|n_j) = \logit^{-1}(\beta_0 + \alpha^{(X_1)}_{j} + \alpha^{(X_2)}_{j} + \alpha^{(X_3)}_{j}),$
    \item Bias variable model: $\mbox{Pr}(y_{i \in j} = 1|n_j) = \logit^{-1}(\beta_0 + \alpha^{(X_1)}_{j} + \alpha^{(X_3)}_{j} + \alpha^{(X_4)}_{j}),$
    \item Nuisance variable model: $\mbox{Pr}(y_{i \in j} = 1|n_j) = \logit^{-1}(\beta_0 + \alpha^{(X_1)}_{j}  + \alpha^{(X_3)}_{j} ),$
\end{itemize}
where $\alpha^{(k)}_j \sim \normal(0,\sigma^{(k)}$), $\sigma^{(k)} \sim t(3,0,2.5)$ and $\beta_0 \sim t(3,0,2.5)$, which are the standard priors in the brms R package \citep{brms1,brms2}.

For each sample iteration and model we perform the following steps:
\begin{enumerate}
    \item Create a population.
    \item Create a sample from this population so that each cell has at least one individual observed, and then according to $\pi_{i \in j}$.
    \item Fit all four models on this sample using brms with default settings.
    \item For each model we calculate
    \begin{enumerate}
        \item The population mean estimate using MRP and score it compared to the truth using CRPS and squared error,
        \item An estimate for each cell and compare to the true $\mbox{Pr}(Y_{i \in j} =1)$ to calculate the cellwise CRPS and squared error as per~\eqref{eqn:mrp_mse} and~\eqref{eqn:mrp_crps}.
    \end{enumerate}
\end{enumerate}

The cellwise score using the population cellwise truth is equivalent, up to a small amount of noise, to the population score; see supplementary materials for visualisation of results). However, in practice the population truth (cellwise or at the aggregate level) is not available. 

One option in this case is to approximate  the population truth with the sample observation, $\mbox{Pr}(y_{i \in j}|n_j,y_{i \in j})$. This leads to

\begin{equation}
\label{eqn:mrp_mse_sample}
    \widehat{\Error}_{\mathrm{MRP}} = \frac{\sum_{j=1}^JN_j(\mbox{Pr}(Y_{i \in j}|y,\hat{\theta})-\mbox{Pr}(y_{i \in j}|n_j, y_j) )}{\sum_{j=1}^J N_j},
\end{equation}
and
\begin{multline}
\label{eqn:mrp_crps_sample}
    \widehat{\CRPS}_{\mathrm{MRP}}(y,n,\hat{\theta},\mathbf{x})  =  \frac{1}{2}\frac{1}{B}\sum_b^B\Bigg(\Bigg\lvert \frac{\sum_{j=1}^JN_j\Big(\mbox{Pr}(Y_{i \in j}|y, \hat{\theta}^b) - \mbox{Pr}(Y_{i \in j}|y, \hat{\theta}'^b)\Big)}{\sum_{j=1}^J N_j}\Bigg\rvert\Bigg)  - \\
    \frac{1}{B}\sum_b^B\Bigg(\Bigg\lvert \frac{\sum_{j=1}^JN_j\Big(\mbox{Pr}(Y_{i \in j}|y, \hat{\theta}^b)- \mbox{Pr}(y_{i \in j}=1|n_j,y_j)\Big)}{\sum_{j=1}^J N_j} \Bigg\rvert\Bigg)    .  
\end{multline}

When there is only one observation per cell, the cellwise probability estimate is either $0$ or $1$ accordingly. Previous research suggests that this will underestimate the magnitude of the score, which we also find (see supplementary materials for a visual demonstration). However, the ordering of models is preserved---namely, the separation between models that contain the bias variable as a predictor and models that contain only the precision variable as a predictor. The standard sum of individual scores does not maintain this ordering (e.g., see \citealp{kuh2022using} for sum of ELPD, and supplementary materials for mean of CRPS and squared error scores), which suggests the scoring rule could be useful for ordering models but not estimating the expected magnitude of error. 

\subsection{Brute-force cross validation}
To resolve the underestimation of the magnitude of scores we consider a cross validation approach. There is a wide and established literature on cross validation for estimating predictions. We considered three potential schemes: $K$-fold cross validation, leave-one-cell-out (LOCO) and leave-one-level-out (LOLO). 

$K$-fold cross validation partitions the sample into $K$ non-overlapping folds. The $k^{th}$ fold is removed to fit the model and score calculated using this fold. To use this approach in with MRP, we would need to adjust the score estimated in this fold to population. To do this we would need at least $K$ observations present in every poststratification matrix cell (one for each fold) which isn't plausible given the size and complexity of survey data and the set of potential adjustment variables. 

Leave-one-cell-out cross validation uses a similar idea, but instead partitions the data so each cell is it's own fold. The $j^{th}$ cell is removed, the model fit and the score estimated using the observations in the removed cell $\mbox{Pr}(y_j=1|n_j)$. In this scenario, only one observation is required per cell (to estimate the probability), and the estimate of population error can be made with our proposed score decomposition. This is the method that we will focus on in this manusucript. 

Despite this,  leave-one-level-out (LOLO) cross validation could be useful specifically in subpopulation estimation (see Section~\ref{sec:sub_pop}). In this instance the data are partitioned by the level of a adjustment variable and a similar process to the other validation techniques applied. In general though, it is not clear which variable to focus on when multiple variables are in the model when assessing a whole population estimate. 

To use LOCO, we propose replacing our estimate $\mbox{Pr}(Y_{i \in j}  = 1|y,\hat{\theta})$ with $\mbox{Pr}(Y_{i \in j}  = 1|y_{i \notin j},\hat{\theta})$, the probability of $y = 1$ in that cell given that it was not used when the model was fit to get
\begin{equation}
\label{eqn:mrp_mse_loco_bruteforce}
    \Error^{(\mathrm{LOCO-CV})} = \frac{\sum_{j=1}^JN_j(\mbox{Pr}(Y_{i \in j}|y_{i \notin j},\hat{\theta})-\mbox{Pr}(y_{i \in j} |n_j, y_j) )}{\sum_{j=1}^J N_j},
\end{equation}
and
\begin{multline}
\label{eqn:mrp_crps_loco_bruteforce}
    \CRPS^{(\mathrm{LOCO-CV})}(y,n,\hat{\theta})  =  \frac{1}{2}\frac{1}{B}\sum_b^B\Bigg(\Bigg\lvert \frac{\sum_{j=1}^JN_j\Big(\mbox{Pr}(Y_{i \in j}|y_{i \notin j}, \hat{\theta}^b) - \mbox{Pr}(Y_{i \in j}|y_{i \notin j}, \hat{\theta}'^b)\Big)}{\sum_{j=1}^J N_j}\Bigg\rvert\Bigg)  - \\
    \frac{1}{B}\sum_b^B\Bigg(\Bigg\lvert \frac{\sum_{j=1}^JN_j\Big(\mbox{Pr}(Y_{i \in j}|y_{i \notin j}, \hat{\theta}^b)- \mbox{Pr}(y_{i \in j}|n_j, y_j)\Big)}{\sum_{j=1}^J N_j} \Bigg\rvert\Bigg).    
\end{multline}
From here on in this manuscript we will only be discussing an aggregate MRP score and so no longer note this for notation readability. 

\begin{figure}[t!]
    \centering
    \includegraphics[width=.9\textwidth]{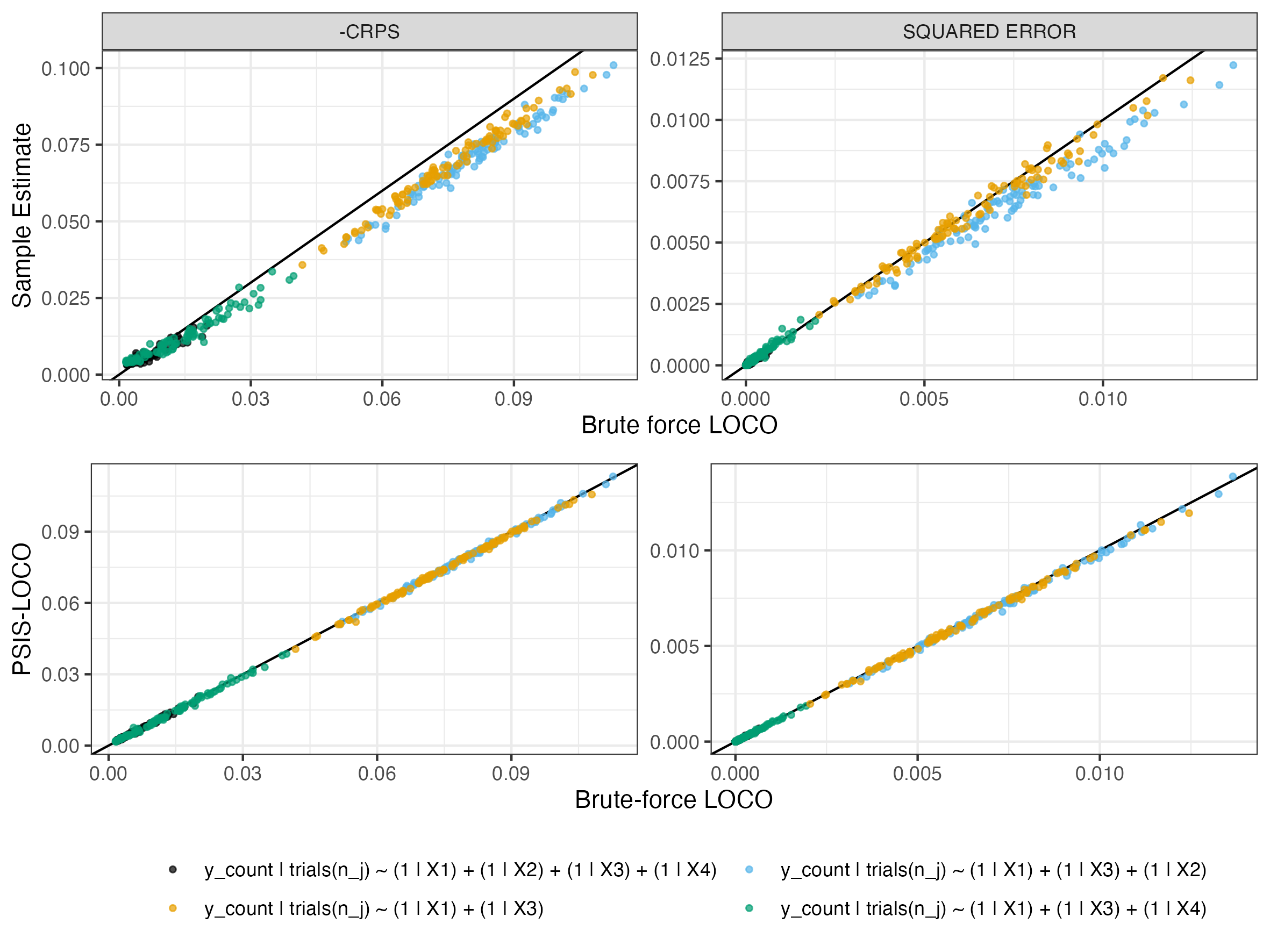}
    \caption{\em Comparison of the brute-force leave-one-cell-out approximation (LOCO, $x$-axis) with a non cross validation scheme ($y$-axis, top row) and approximate cross validation scheme ($y$-axis, bottom row). Colour of point represents model, points within this represent different simulation iterations. The black line represents an identity line. We plot (-)CRPS(left panels) and squared error (right panels). }
    \label{fig:bruteforce_samplescore}
\end{figure}

\begin{figure}[t!]
    \centering
    \includegraphics[width=.9\textwidth]{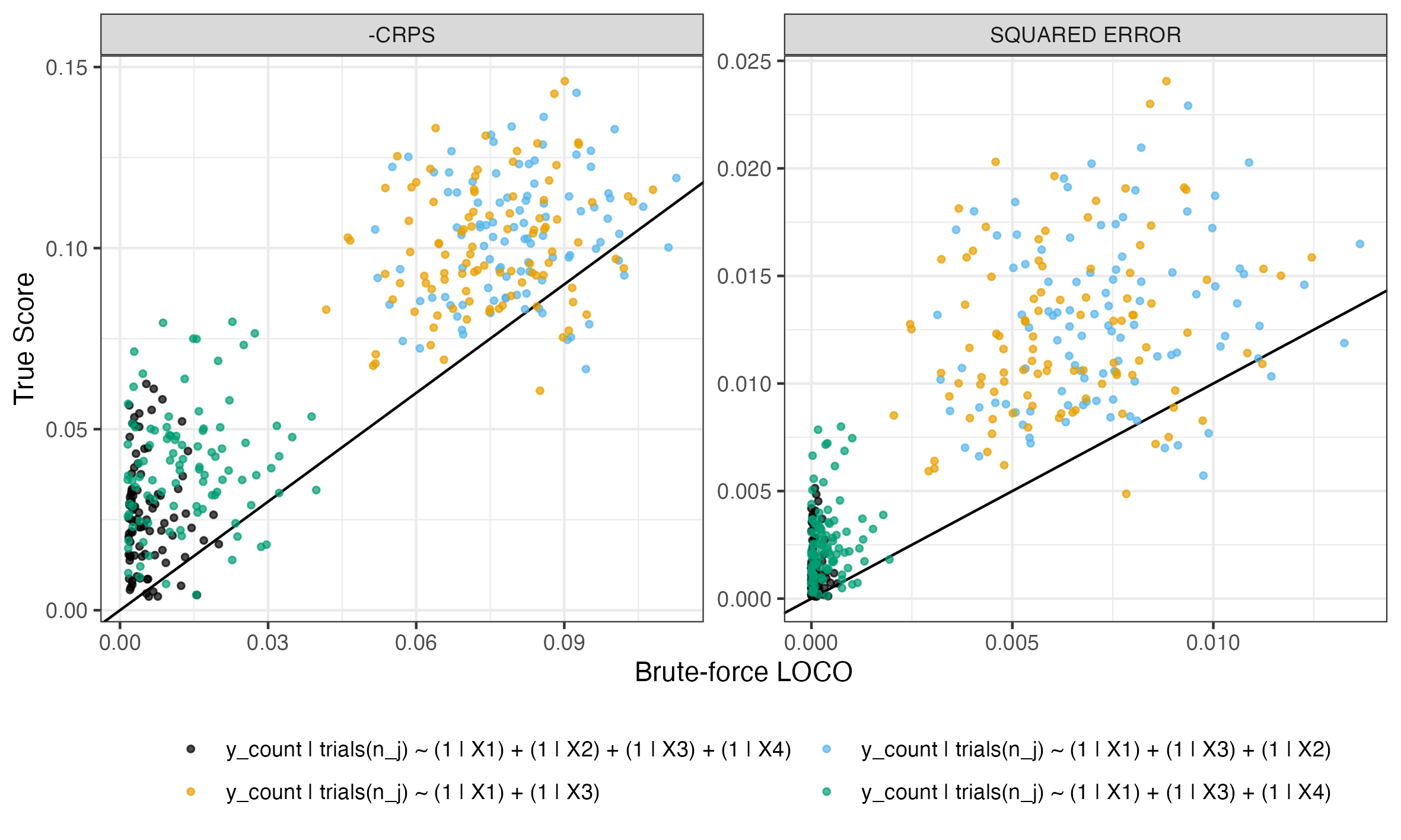}
    \caption{\em Comparison of a direct score of the MRP estimate (using known population truth, $y$-axis) against an indirect score of the MRP estimate (using a brute-force leave-one-cell-out approach, $x$-axis). Colour of point represents the different models fitted, points within this different simulation iterations. The black line represents an identity line. We plot squared error and the negative of CRPS, so higher can be interpreted as a worse model. These results demonstrate that using LOCO still underestimates the model score. }
    \label{fig:bruteforce_truth}
\end{figure}

When the model is refit excluding each cell (leading to $J+1$ model fits), we call this a brute-force implementation of leave-one-cell-out (LOCO). It is not efficient in terms of time and computational resources, but it does allow us to directly investigate whether the underestimation impact of cross validation on the magnitude of score estimate. Figure~\ref{fig:bruteforce_samplescore} shows that the score estimated with a brute-force LOCO approach is indeed slightly larger than that estimated using the full sample to model and estimate the score. However, this is still a considerable underestimate for the true score, as shown by Figure~\ref{fig:bruteforce_truth}.

Despite the underestimate of the magnitude of the score, we believe that this approach is still useful to practitioners who wish to evaluate the relative efficacy of different models when estimating the population mean. Unlike previous attempts, ordering is maintained. We advise that the magnitude of the score be interpreted with caution. 

\subsection{Approximate cross validation }
\label{sec:psis_approx}
The magnitude of the score is not the only challenge for a brute-force leave-one-cell-out approach. A practical issue is the sheer amount of computational required for a brute-force approach. For example, in the first iteration of the previously described simulation study, the posterior inference had to be rerun $294$ times (once for each cell plus one for full data), and this example is a relatively small toy problem in terms of the size of the poststratification matrix. Many MRP examples involve poststratifying to an area level variable like postcode, which might have $100$ levels \citep{machalek2022serological}. In this section, we apply an approximate LOO method based on Pareto smoothed importance sampling \citep[PSIS; ][]{vehtari2017practical,vehtari2022pareto}. 

PSIS-LOO (equation (10) of \citealp{vehtari2017practical}) estimates the sum expected log individual predictive density by
\begin{equation*}
    \widehat{\elpd}_{\mathrm{PSIS-LOO}} = \sum_{i=1}^n \log\Bigg(\frac{\sum_{b=1}^Bw_i^b\mbox{Pr}(y_i|\theta^b)}{\sum_{b=1}^Bw_i^b}\Bigg),
\end{equation*}
where $w_i^b$ is the Pareto smoothed importance weight for the $b^{th}$ posterior draw \citep[denoted $s$ by ][]{vehtari2017practical} and the $i^{th}$ individual. We modify this for our purposes by using other scoring metrics. We begin with substituting the error in each cell as before, with for shorthand we denote $\epsilon_j^b =  \mbox{Pr}(Y_{i \in j} |y_{i \notin j},\hat{\theta}^b)-\mbox{Pr}(Y_{i \in j} |n_j,y_j)$. For each cell $j$ we can compute Pareto smoothed weights $w_j^b$ for each posterior sample $b$ to estimate $\epsilon_j^b$ as outlined by \cite{vehtari2017practical}. Using these, we can approximate $\epsilon_j^b$ and substitute into \eqref{eqn:mrp_mse_loco_bruteforce} as
\begin{equation}
\label{eqn:mrp_mse_loco_psis}
    \widehat{\Error}^{(\mathrm{LOCO-CV})} = \frac{1}{N}\sum_{j=1}^J N_j\frac{\sum_{b=1}^B w_j^b \epsilon_j^b}{\sum_{b=1}^Bw_j^b}
\end{equation}
and thus estimate the squared error of the MRP estimate using the full model fit. 

It is slightly more complicated to achieve this with the CRPS, but we begin by denoting the $g(y_{i \notin j}, \hat{\theta}^b,\hat{\theta}'^b) = \mbox{Pr}(Y_{i \in j}|y_{i \notin j}, \hat{\theta}^b) - \mbox{Pr}(Y_{i \in j}|y_{i \notin j}, \hat{\theta}'^b)$ and $h(y_{i \notin j},y_j,n_j,\hat{\theta}^b) = \mbox{Pr}(Y_{i \in j}|y_{i \notin j}, \hat{\theta}^b)- \mbox{Pr}(y_{i \in j}|n_j, y_j)$. This means we can write \eqref{eqn:mrp_crps_loco_bruteforce} more concisely as
\begin{equation*}
\label{eqn:mrp_crps_loco_bruteforce_simple}
    \begin{split}
    \CRPS^{(\mathrm{LOCO-CV})}(y,n, \hat{\theta},\hat{\theta}')  =  \frac{1}{2}\frac{1}{B}\sum_{b=1}^B\Bigg(\Bigg\lvert \frac{\sum_{j=1}^JN_jg(y_{i \notin j},\hat{\theta}^b,\hat{\theta}'^b)}{\sum_{j=1}^J N_j}\Bigg\rvert\Bigg)  - \\ 
    \frac{1}{B}\sum_{b=1}^B\Bigg(\Bigg\lvert \frac{\sum_{j=1}^JN_jh(y_{i \notin j},y_j,n_j,\hat{\theta}^b)}{\sum_{j=1}^J N_j} \Bigg\rvert\Bigg).
    \end{split}
\end{equation*}

While it is possible to simply take an importance-weighted sum for the squared error, CRPS doesn't have an amenable form. To apply the PSIS weights, which are the $b^{th}$ weight for the $j^{th}$ cell, we do importance resampling of the posterior draws for each cell using a stratified resampling algorithm \citep{kitagawa1996monte} using the $B$ PSIS weights for each cell. This procedure is implemented in the posterior R package \citep{posterior}. This produces $B$ resampled posterior draws $\hat{\theta}_{RW}$, which we permute to calculate PSIS reweighted $\hat{\theta}_{RW}'$, and thus $g_j'(\hat{\theta}_{RW}^b,\hat{\theta}_{RW}'^b) = \mbox{Pr}(Y_{i \in j}  |\hat{\theta}_{RW}^b) - \mbox{Pr}(Y_{i \in j}  |\hat{\theta}_{RW}'^b)$ and $h_j'(y_j,n_j,\hat{\theta}_{RW}^b) = \mbox{Pr}(Y_{ i\in j} |\hat{\theta}_{RW}^b) - \mbox{Pr}(Y_{i\in j}|n_j,y_j)$, to get
\begin{multline}
\label{eqn:mrp_crps_loco_psis}
    \widehat{\CRPS}^{(\mathrm{LOCO-CV})}(y, n, \hat{\theta}_{RW},\hat{\theta}_{RW}')  =  \\ \frac{1}{BN}\Bigg(\frac{1}{2}\sum_{b=1}^B\Bigg(\Bigg\lvert \sum_{j=1}^JN_jg_j(\hat{\theta}_{RW}^b,\hat{\theta}_{RW}'^b)\Bigg\rvert  - 
    \Bigg\lvert \sum_{j=1}^Jh_j(y_j,n_j,\hat{\theta}_{RW}^b) \Bigg\rvert\Bigg)\Bigg).
\end{multline}

The second row of Figure~\ref{fig:bruteforce_samplescore} shows that the approximate PSIS-LOCO scores are approximately equivalent to the brute-force LOCO scores.

Through these simulations, we demonstrate the efficacy of our proposed cellwise scoring in a leave-one-cell-out approach. Whilst the model ordering is well preserved (which is desirable as other approaches do not preserve this), the scores as estimated with the sample are  more favourable to the quality of the estimate than they should be. Using a leave-one-cell-out approach only marginally improves on this, but can be well approximated with a fast approximation such as PSIS-LOCO. Although it is frustrating to underestimate to this degree, we feel the preservation of model ordering still makes this a useful tool, albeit one where the scale must be interpreted carefully. 

When we followed the traditional approach of comparing the mean of cell scores to the true mean of cell scores, the leave-one-cell-out approach appeared to slightly over estimate the magnitude of score (see supplementary), suggesting that the observed underestimation is a feature of our particular score decomposition.

\section{Differentiating between population and subpopulation goodness}
\label{sec:sub_pop}

One of the benefits of MRP is that the same model can be used for both population and subpopulation estimates. To demonstrate how, we follow \cite{kuh2022using} to use set notation to describe the set of cells that form the subpopulation as $\mathbb{S}$. The total number of cells in this set is $S$ and any particular cell as $s$. In the case of the full population, $\mathbb{S}$ is all cells in the population, $S = J$ and $s = j$. Thus we can express a population weighted estimate for any subpopulation (small area) in the population as

\begin{equation}
\label{eqn:est_y_param_sae}
    \E_{\mathrm{S}}(Y|y,\hat{\theta}) = \frac{\sum_{s=1}^SN_s\mbox{Pr}(Y_{i \in s}|y,\hat{\theta})}{\sum_{s=1}^SN_s}.
\end{equation}

One benefit of the leave-one-cell-out estimation technique described in the previous section is that it is also constructed within each cell. This means that we can apply both the CRPS score and squared-error score described previously by focusing on the set of cells that describe a particular subpopulation. In the previous section we demonstrated the accordance of the PSIS-LOCO approximation to brute-force LOCO, and so in this section we use only the PSIS-LOCO for computational efficacy. 

In this study we add two more models to our comparison set to add great distinction for the $X_1$ and $X_3$ subpopulation estimates:

\begin{itemize} 
    \item $X_1~\mathrm{only}$ : $\mbox{Pr}(y_{i \in j} = 1|n_j) = \logit^{-1}(\beta_0 + \alpha^{(X_1)}_{i \in j}),$
    \item $X_3~\mathrm{only}$ : $\mbox{Pr}(y_{i \in j} = 1|n_j) = \logit^{-1}(\beta_0 + \alpha^{(X_3)}_{i \in j}),$
\end{itemize}

We can modify \eqref{eqn:mrp_mse_loco_psis} to estimate the subpopulation score with 
\begin{equation}
\label{eqn:mrp_mse_loco_psis_sae}
    \widehat{\Error}_S^{(\mathrm{LOCO-CV})} = \frac{1}{\sum_{s=1}^SN_s}\sum_{s=1}^S N_s\frac{\sum_{b=1}^B w_s^b \epsilon_s^b}{\sum_{b=1}^Bw_s^b},
\end{equation}
by simply limiting to the relevant subpopulation rather than the full population. Similarly we can estimate the CRPS for the subpopulaton by modifying \eqref{eqn:mrp_crps_loco_psis} to similarly focus on the relevant subpopulation:
\begin{multline}
\label{eqn:mrp_crps_loco_psis_rw}
    \widehat{\CRPS}_S^{(\mathrm{LOCO-CV})}(y,n, \hat{\theta}_{RW},\hat{\theta}'_{RW})  = \\ \frac{1}{BN}\sum_{b=1}^B\Bigg(\frac{1}{2}\Bigg\lvert \sum_{s=1}^SN_sg_s(\hat{\theta}_{RW}^b,\hat{\theta}_{RW}'^b)\Bigg\rvert  - 
    \Bigg\lvert \sum_{s=1}^SN_sh_s(y_s,n_s,\hat{\theta}_{RW}^b) \Bigg\rvert\Bigg),
\end{multline}
using the same importance resampling technique as described previously. 

It is common for multiple subpopulations to be estimated, most often all levels of a variable (e.g., geographic regions). This means there is interest in scoring the goodness of estimates for all levels of a particular variable rather than one specific level. To describe this we use $l$ to denote a particular level of the $k^{th}$ variable, with the total number of levels in this variable $L$. Where we it is necessary to identify a specific variable we denote with a superscript $(k)$ but otherwise assume this is clear from the context for simplicity. For a particular score, it is possible to plot and consider $\Score_l$, for every $l$, but it is also possible to consider the average of scores for every level:
\begin{equation}
    \label{eqn:sae_averagescore}
    \Score^{(k)} = \frac{\sum_{l=1}^{L^{(k)}}\Score_l}{L^{(k)}},
\end{equation}
which in the case of the score being squared error, corresponds to the mean squared error over levels of $k$. 

\begin{figure}[ht!]
    \centering
    \includegraphics[width=\textwidth]{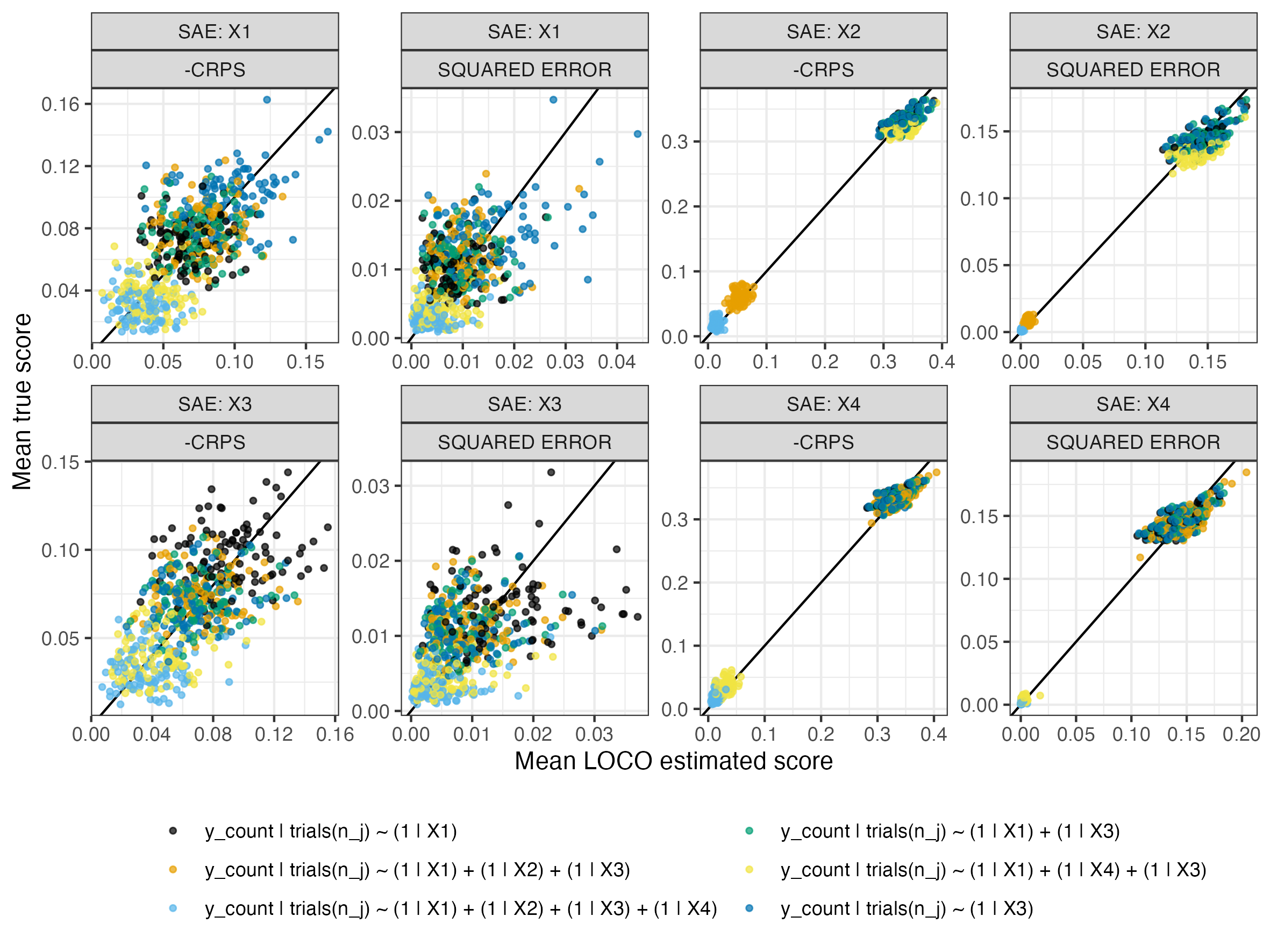}
    \caption{\em Comparison of the mean of true scores across levels ($y$-axis) against the mean of estimated scores ($x$-axis, alternating vertical facets) for each level in each variable (horizontal and vertical facets). Colour of the point represents the different models fitted, whilst different points within this represent different simulation iterations. The black line represents an identity line. These results can be separated into two classes, those estimating levels of X2 and X4 (strongly predictive of outcome) and those estimating levels of X1 and X3 (weakly predictive of outcome). }
    \label{fig:sae_mean}
\end{figure}

Figure~\ref{fig:sae_mean} shows a comparison between the mean over levels of the estimated PSIS-LOCO score for each variable and the true mean over levels. Focusing on the two variables that are strongly predictive of the outcome ($X_2$ and $X_4$; final two columns) we can cleanly distinguish between models with a low mean true score and models with a high mean true score. This was a demonstrated challenge in \cite{kuh2022using} that these new metrics overcome. Of particular interest for the practitioner is that the models with the lowest true score differ for the different variables, which again reinforces the concept proposed by \cite{kuh2022using} that the best model differs based on the overall aim. 

Turning our attention to the variables that are less predictive of the outcome ($X_1$ and $X_3$; first two columns), we see less clear delineation between good and bad models. However, we still retain the respective ordering despite the noisiness of estimates. The respective score of each level can also be plotted individually; see supplementary materials.

\section{Reference validation score}

To use our scoring decomposition we require at least one observation in each cell. However, in many real-world cases not all cells will be observed in the sample. To overcome this, we follow the approach of \cite{vehtari2012survey} and substitute the observed cellwise probability with an estimated probability using a reference model denoted $M_*$. To investigate this, we first consider a full reference validation score where we make this substitution for every cell (Section~\ref{sec:allobs}), which allows us to compare to cross validation. Then we describe a new simulation scenario where not all cells are observed. We first demonstrate PSIS LOCO reference validation, and then propose a LOCO combined validation approach in which we combine cross validation and reference validation. As a consequence of this, we describe a way of validating the chosen reference model. 

\subsection{All cells observed}
\label{sec:allobs}
The reference validation technique follows the approach of  \cite{vehtari2012survey}. We denote the reference model as $M_*$, and compare to a candidate model, $M_c$, with the overall aim to evaluate to rank the $C$ candidate models  and identify candidate models that perform similarly to the reference model. The reference model can be difficult to propose \citep[as discussed by][]{vehtari2012survey}, but in an MRP context there is often need to identify if a smaller or simpler model would suffice . In this paper the model with $X_4$ could be considered of similar efficacy to a model with all predictors. 


We can apply a reference validation approach by substituting the reference model estimate for the population truth in \eqref{eqn:mrp_mse} and~\eqref{eqn:mrp_crps}. In Equation (30) of \cite{vehtari2012survey}, the expected loss function under the actual belief model $M_*$ is decomposed in the variance for a predicted value given $M_*$ and the squared difference in expectations between the reference model $M_*$ and the proposed model $M_c$. As mentioned by \cite{vehtari2012survey} the model minimising the squared loss is that which the predicted expectation is closest to that with the reference model. Incorporating this into~\eqref{eqn:mrp_mse} in an MRP context, we would aim to minimise the square of
\begin{equation*}
    \Error^{(\mathrm{REF})}( M_c, M_*) = E_P(Y|M_c, \theta_c) - E_P(Y|M_*, \theta_*),
\end{equation*}
which can be expanded in a similar form as in previous sections to 
\begin{equation}
\label{eqn:reference_mrp_mse}
    \Error^{(\mathrm{REF})}( M_c, M_*) = \frac{\sum_{j=1}^JN_j(\mbox{Pr}(Y_{i \in j}|\theta_c) - \mbox{Pr}(Y_{i \in j} |\theta_*))}{\sum_{j=1}^J N_j},
\end{equation}

with $\mbox{Pr}(Y_{i \in j}|\theta_c)$ shorthand for $\mbox{Pr}(Y_{i \in j}=1|M_c, \theta_c, \textbf{x}, y)$.

Similarly we use Lemma (17) of \cite{szekely2005new} to measure the difference between the predictive distribution of $M_*$ and $M_c$. This lemma is also used by \cite{gneiting2007strictly} when discussing CRPS:
\begin{equation*}
     \mathrm{CRPS} = \int_{-\infty}^\infty(G(t)-F(t))^2dt = \frac{1}{2}E |X - X'| + \frac{1}{2}E |Y - Y'| - E |X - Y|,
\end{equation*}
where $F$ and $G$ are the cumulative distribution functions for the reference and candidate models, respectively, and so we use $\mbox{Pr}(Y|\theta_*)$ as $Y$ and $\mbox{Pr}(Y|\theta_c)$ as $X$. Plugging these in and simplifying as in previous sections, we get
\begin{align*}
      \mathrm{CRPS}^{(\mathrm{REF})}( M_c, M_*) = \frac{1}{BN} \Bigg(\frac{1}{2} \sum_{b=1}^B\Bigg|\sum_{j=1}^JN_j\bigg(\mbox{Pr}(Y_{i\in j}|\theta_c^b)-\mbox{Pr}(Y_{i\in j}|\theta_c'^b\bigg)\Bigg| \,+\\
      \qquad \frac{1}{2} \sum_{b=1}^B\Bigg|\sum_{j=1}^JN_j\bigg(\mbox{Pr}(Y_{i\in j}|\theta_*^b)-\mbox{Pr}(Y_{i\in j}|\theta_*'^b\bigg)\Bigg|\, - \\
        \sum_{b=1}^B\Bigg|\sum_{j=1}^JN_j\bigg(\mbox{Pr}(Y_{i\in j}|\theta_c^b)-\mbox{Pr}(Y_{i\in j}|\theta_*'^b\bigg)\Bigg|\,\Bigg) .   
\end{align*}
We can implement the LOCO estimate of $\mbox{Pr}(Y_{i\in j}|\theta_c)$, $\mbox{Pr}(y_{i\in j}|\theta_c, y_{\notin j})$ to avoid overfitting and more accurately capture the predictive power for unseen observations. For squared error, we use the square of
\begin{equation}
\label{eqn:reference_mrp_mse_loco}
\begin{split}
    \Error^{(\mathrm{LOCO-REF})}(M_c, M_*) = \frac{\sum_{j=1}^JN_j(\mbox{Pr}\left(Y_{i \in j}|M_c, y_{i \notin j}\right) - \mbox{Pr}\left(Y_{i \in j}|M_*, y_{i \notin j}\right))}{N},
\end{split}
\end{equation}
and for CRPS we have 
\begin{align*}
      \mathrm{CRPS}^{(\mathrm{LOCO-REF})}( M_c, M_*) = \frac{1}{BN} \sum_{b=1}^B\Bigg(\frac{1}{2} \Bigg|\sum_{j=1}^JN_j\bigg(\mbox{Pr}(Y_{i\in j}|\theta_c^b, y_{i\notin j})-\mbox{Pr}(Y_{i\in j}|\theta_c'^b,y_{i\notin j}\bigg)\Bigg| \,+\\
      \qquad \frac{1}{2}\Bigg|\sum_{j=1}^JN_j\bigg(\mbox{Pr}(Y_{i\in j}|\theta_*^b,y_{i\notin j})-\mbox{Pr}(Y_{i\in j}|\theta_*'^b,y_{i\notin j}\bigg)\Bigg|\, - \\
        \Bigg|\sum_{j=1}^JN_j\bigg(\mbox{Pr}(Y_{i\in j}|\theta_c^b,y_{i\notin j})-\mbox{Pr}(Y_{i\in j}|\theta_*'^b,y_{i\notin j}\bigg)\Bigg|\,\Bigg) .   
\end{align*}

The PSIS approximation of the LOCO reference model estimate for each cell can be substituted. For the error score, we simply use an importance-weighted estimate for each cell:
\begin{multline}
\label{eqn:reference_mrp_mse_loco_psis}    \widehat{\Error}^{(\mathrm{LOCO-REF})}(M_c, M_*) = \\
\frac{1}{N}\sum_{j=1}^JN_j\bigg(\frac{\sum_{b=1}^B w_{c,j}^b \mbox{Pr}(Y_{i \in j}|\hat{\theta}^b_{c})}{\sum_{b=1}^Bw_{c,j}^b} - 
    \frac{\sum_{b=1}^B w_{*,j}^b \mbox{Pr}(Y_{i \in j}|\hat{\theta}^b_{*})}{\sum_{b=1}^Bw_{*,j}^b}\bigg).
\end{multline}

\begin{figure}[t!]
    \centering
    \includegraphics[width=.9\textwidth]{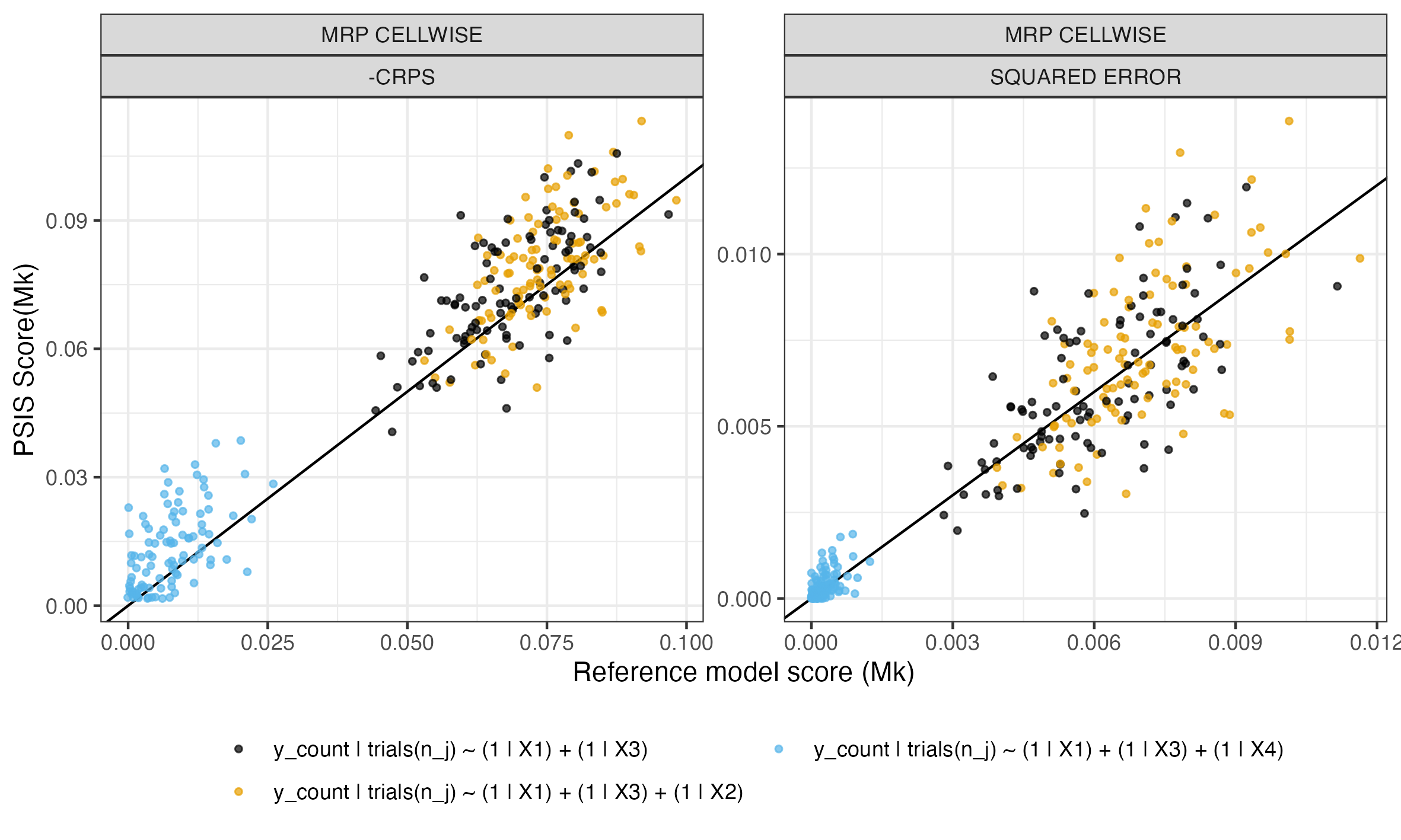}
    \caption{\em Comparison of the candidate model LOCO cross validation score (when estimated using Pareto smoothed importance sampling, $y$-axis) against the LOCO reference validation approach ($x$-axis). Colour of point represents the different candidate models, whilst different points within this represent different simulation iterations. The black line represents an identity line. The strong relationship indicates that when the reference model is well chosen, this closely approximates the PSIS-LOCO score.}
    \label{fig:reference_vs_psisloco}
\end{figure}

For the LOCO reference validation CRPS score we can use the importance-resampled posteriors for reference model $M_*$, $\hat{\theta}_{*,RW}$ and candidate model $M_c$, $\hat{\theta}_{c,RW}$. We then modify as follows: $g_{j,\mathrm{REF}}'(\hat{\theta}_{c,RW}^b,\hat{\theta}_{c,RW}'^b) = \mbox{Pr}(Y_{i \in j}|\hat{\theta}_{c,RW}) - \mbox{Pr}(Y_{i \in j}|\hat{\theta}_{c,RW}'^b)$ and $h_{j,\mathrm{REF}}'(\hat{\theta}_{c,RW}^b,\hat{\theta}_{*,RW}^b) = \mbox{Pr}(Y_{i \in j}| \hat{\theta}_{c, RW}^b)- \mbox{Pr}(Y_{i \in j}|\hat{\theta}_{*,RW}^b)$.
\begin{eqnarray}
\nonumber
\widehat{\CRPS}^{(\mathrm{LOCO-REF})}( M_c, M_*) = 
      \frac{1}{BN}\sum_{b=1}^B\Bigg(\frac{1}{2}\Bigg|\sum_{j=1}^JN_jg'_{j,\mathrm{REF}}(\hat{\theta}_{c,RW}^b,\hat{\theta}_{c,RW}'^b)\Bigg|\, +\\
 \frac{1}{2}\Bigg|\sum_{j=1}^JN_jg'_{j,\mathrm{REF}}(\hat{\theta}_{*,RW}^b,\hat{\theta}_{*,RW}'^b)\Bigg|\,- 
     \Bigg|\sum_{j=1}^JN_jh'_{j,\mathrm{REF}}(\hat{\theta}_{c,RW}^b,\hat{\theta}_{*,RW}^b)\Bigg|\,\Bigg). 
\end{eqnarray}

To demonstrate the efficacy of the PSIS LOCO reference validation approach, we let the full model with all covariates be $M_*$ and the candidate models be the model without the precision variable, the model without the bias variable, and the model without either. For each we use a PSIS approximation of the predictive error of cell $j$ and apply this to our previous example with all cells observed. Figure~\ref{fig:reference_vs_psisloco} demonstrates that the PSIS LOCO reference validation approach and the PSIS LOCO cross validation approach described in Section~\ref{sec:estimating_goodness} produce similar estimates when the appropriate reference model is chosen. 

\subsection{Not all cells observed}
\label{sec:not_all_obs}
Previously, we demonstrated the equivalence of a reference validation approach to a cross validation approach when all cells are observed. Here, we relax the constraint that all cells are observed. We contribute three components. Firstly we detail how to use the reference validation approach in this case. Secondly, we propose a technique to validate the reference model. Thirdly, we propose a combined validation method using LOCO cross validation and LOCO reference validation.

\subsubsection{Reference validation score}
\label{sec:not_all_ref}
We begin by partitioning the population into two sets. The first set, $\mathbb{V}$, is the set of cells where at least one observation in each is observed in the sample. The total number of cells in this set is $V$ and any particular cell in this set is $v$. The second set, $U$, is the set of cells that do not have any observations in the sample. The total number of cells in this set is $U$ and any particular cell in this set is $u$. For clarity note that $J = V + U$.

When cells are observed, we could use a cross validation or reference validation approach to assess predictive power. When cells are not observed, we can only use a reference validation approach. To minimise the reliance on correctly specifying the reference model, we could combine cross validation for observed cells and reference validation for unobserved cells. For simplicity, we write the Pareto smoothed LOCO approximation as
\begin{equation*}
    \mbox{Pr}(Y_{i \in j}|\hat{\theta}^b_{c}, w_{c,j}) =\frac{\sum_{b=1}^B w_{c,j}^b \mbox{Pr}(Y_{i \in j}|\hat{\theta}^b_{c} )}{\sum_{b=1}^Bw_{c,j}^b}.
\end{equation*}
and our modified equation as
\begin{multline}
 \Error^{(\mathrm{REF})}(M_c, M_*) = 
\frac{1}{N}\Bigg(\sum_{v=1}^{V}\!\!N_{v}\bigg( \mbox{Pr}(Y_{i \in v}|\hat{\theta}^b_{c},w_{c,v})- 
    \mbox{Pr}(Y_{i \in v}|\hat{\theta}^b_{*},w_{*,v})\bigg) + \\
\label{eqn:reference_mrp_se_loco_psis_partial}
    \sum_{u=1}^{V}N_{u}\bigg(
    \mbox{Pr}(Y_{i \in u}|\hat{\theta}_{c}) -  \mbox{Pr}(Y_{i \in u}|\hat{\theta}_{*}) 
    \Bigg).
\end{multline}
Each component of CRPS can similarly be decomposed into an approximate LOCO reference validation for the cells with observations and approximate LOCO reference validation for the cells with no sample observations. We denote $g_{j}^{(\mathrm{REF})}(.)$ and $h_{j}^{(\mathrm{REF})}(.)$ as the corresponding $g_j(.)$ and $h_j(.)$ for the reference model approach.  

\begin{multline}
\widehat{\CRPS}^{(\mathrm{REF})}( M_c, M_*) = \\
\frac{1}{BN}\sum_{b=1}^B \Bigg(\frac{1}{2}\Bigg|\sum_{v=1}^{V}N_{v}g_{v}^{(\mathrm{REF})}(\hat{\theta}_{c}^b,\hat{\theta}_{c}'^b) + 
\sum_{u=1}^{U}N_{u}g_{u}^{(\mathrm{REF})}(\hat{\theta}_{c}^b,\hat{\theta}_{c}'^b)\Bigg|\, +
\\
+ \ \frac{1}{2} \Bigg|\sum_{v=1}^{V}N_{v}g_{v}^{(\mathrm{REF})}(\hat{\theta}_{*}^b,\hat{\theta}_{*}'^b) + 
\sum_{u=1}^{U}N_{u}g_{u}^{(\mathrm{REF})}(\hat{\theta}_{*}^b,\hat{\theta}_{*}'^b)\Bigg|\, -\\
 \ \Bigg| \sum_{v=1}^{V}N_{v}h_{v}^{(\mathrm{REF})}(\hat{\theta}_{c}^b,\hat{\theta}_{*}^b) + 
\sum_{u=1}^{U}N_{u}h_{u}^{(\mathrm{REF})}(\hat{\theta}_{c}^b,\hat{\theta}_{*}^b)\Bigg|\,\Bigg).   
\label{eqn:reference_mrp_crps_loco_psis_partial}  
\end{multline}

\subsubsection{Combined validation score}

The challenge with the method described in Section~\ref{sec:not_all_ref} is that it doesn't use the information that we have observed in the sample to test model goodness. In this section we utilise this information in two ways. Firstly we propose a technique to provide partial validation for the reference model. Secondly, we propose a modification of \eqref{eqn:reference_mrp_se_loco_psis_partial} and~\eqref{eqn:reference_mrp_crps_loco_psis_partial} to use LOCO reference   validation when the cells are not observed and LOCO cross validation when they are. 

We begin by focusing on the portion of the population that is observed. We  calculate the PSIS LOCO reference validation score and the PSIS LOCO cross validation score by treating the observed cells as a subpopulation; see Section~\ref{sec:sub_pop}. By comparing the resultant model ordering we can identify whether the two approaches suggest similar ordering for the subpopulation where the sample has an observation in each cell. 

To demonstrate this, we extend the simulations in Section \ref{sec:aggregate_scoring} to relax the assumption that every cell needs to be observed in the sample. Instead we simply constrain that every level of a variable needs to be observed in the sample when taking a sample from the population. With this constraint relaxed, our simulated samples contained $29\%$ of the cells in the population (minimum $25\%$ and maximum $33\%$).

\begin{figure}[p]
    \centering
    \includegraphics[width=.9\textwidth]{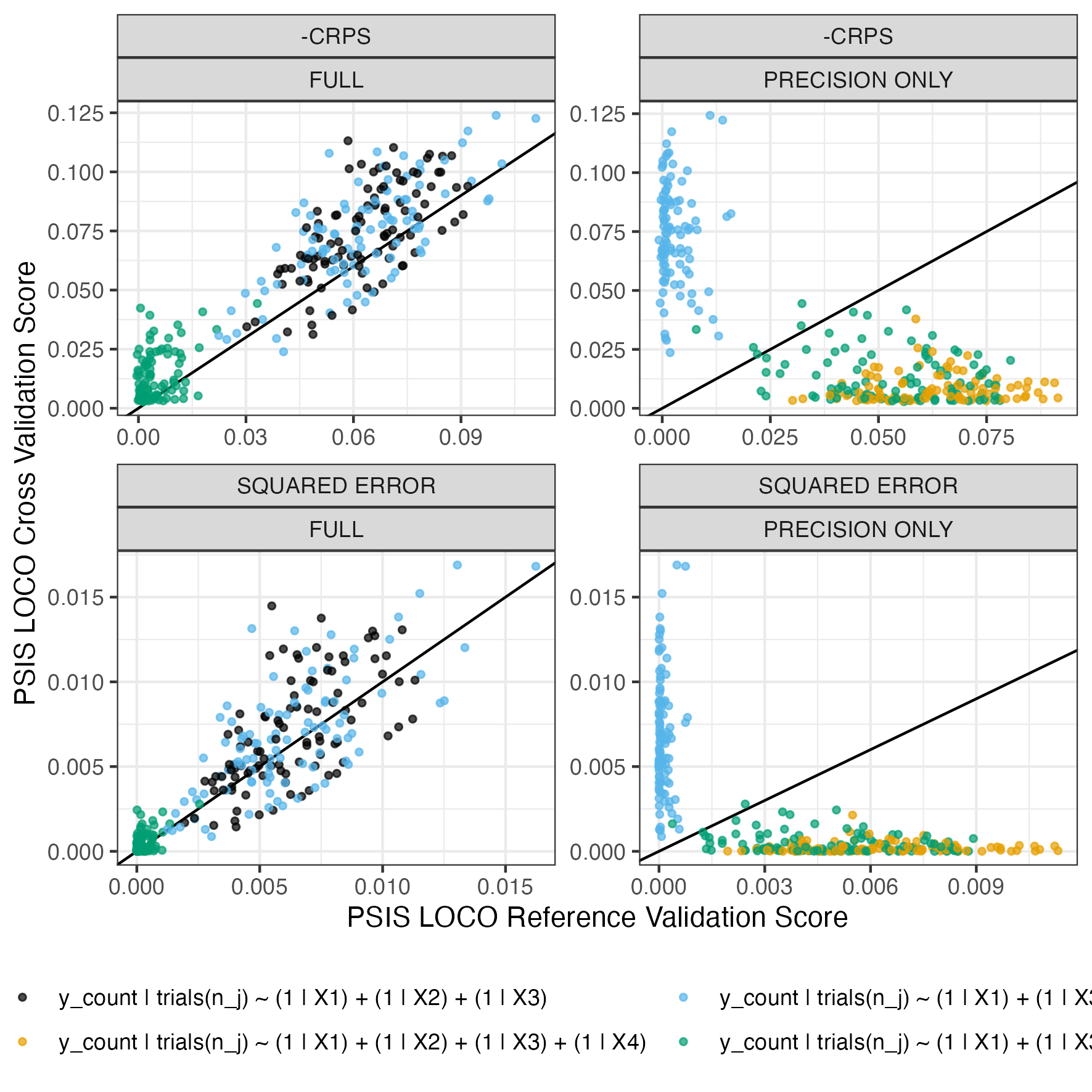}
    \caption{\em Comparison of the PSIS LOCO cross validation score ($x$-axis) against the PSIS LOCO reference validation score ($y$-axis). Colour of point represents the different candidate models, whilst different points within this represent different simulation iterations. The top row represents the CRPS scores when the reference model is the full model (left panel) and the precision model (right panel). The bottom represents the same for the squared error.}
    \label{fig:reference_vs_psisloco_partial}
\end{figure}

Figure~\ref{fig:reference_vs_psisloco_partial} plots the PSIS LOCO cross validation scores scores against the reference model scores. The first column uses the full model as the reference model, whilst the second column uses the precision model as the reference model. When the reference model is chosen appropriately (the full model) the ordering of scores align, but when an inappropriate reference model is chosen (the precision model), they do not. 

However, this only evaluates part of the population and should not be used to score the models. Instead we propose an approach that combines the reference validation score and the cross validation score. We use the reference validation approach when we don't have an observation in a cell, and the cross validation approach when we do. For squared error we modify  \eqref{eqn:reference_mrp_se_loco_psis_partial} to 
\begin{eqnarray}
\nonumber
 &&   \widehat{\Error}^{(\mathrm{CB})}(M_c, M_*) = \frac{1}{N}\bigg(\sum_{v=1}^VN_v
\frac{\sum_{b=1}^B w_v^b \epsilon_v^b}{\sum_{b=1}^Bw_v^b}\ +
 \label{eqn:reference_mrp_se_loco_psis_combined} \sum_{u=1}^{U}\!\!N_u\!\bigg(
    \mbox{Pr}(Y_{i \in u}|\hat{\theta}_{c}) - 
 \mbox{Pr}(Y_{i \in u}|\hat{\theta}_{*}) 
    \bigg)\!\!\bigg)\!\!,
\end{eqnarray}
where $w_{\mathrm{obs}}^b$ and $\epsilon_{\mathrm{obs}}^b$ are as defined in Section~\ref{sec:psis_approx}. 
Similarly we take \eqref{eqn:reference_mrp_crps_loco_psis_partial} and modify to use $g(.)$ and $h(.)$ as defined in Section~\ref{sec:psis_approx} where the cell is observed:
\begin{multline}
\label{eqn:reference_mrp_crps_loco_psis_combined}
\widehat{\CRPS}^{(\mathrm{CB})}(M_c, M_*) = \\
\frac{1}{BN}\sum_{b=1}^B \Bigg( \frac{1}{2}\Bigg|\sum_{v=1}^{V}N_{v}g_{v}(\hat{\theta}_{c,RW}^b,\hat{\theta}_{c,RW}'^b) + 
\sum_{u=1}^{U}N_{u}g_{u}^{(\mathrm{REF})}(\hat{\theta}_{c,RW}^b,\hat{\theta}_{c,RW}'^b)\Bigg|\, +
\\
+ \ \frac{1}{2} \Bigg|\sum_{v=1}^{V}N_{v}g_{v}(\hat{\theta}_{*,RW}^b,\hat{\theta}_{*,RW}'^b) + 
\sum_{u=1}^{U}N_{u}g_{u}^{(\mathrm{REF})}(\hat{\theta}_{*}^b,\hat{\theta}_{*}'^b)\Bigg|\, -\left | 
\sum_{u=1}^{U}N_{u}h_{u}^{(\mathrm{REF})}(\hat{\theta}_{c}^b,\hat{\theta}_{*}^b) \right| 
\Bigg).   
\end{multline}
There is one less term on the final line as there is no distribution for these estimates in the observed cell case. 

\begin{figure}[t!]
    \centering
    \includegraphics[width=.9\textwidth]{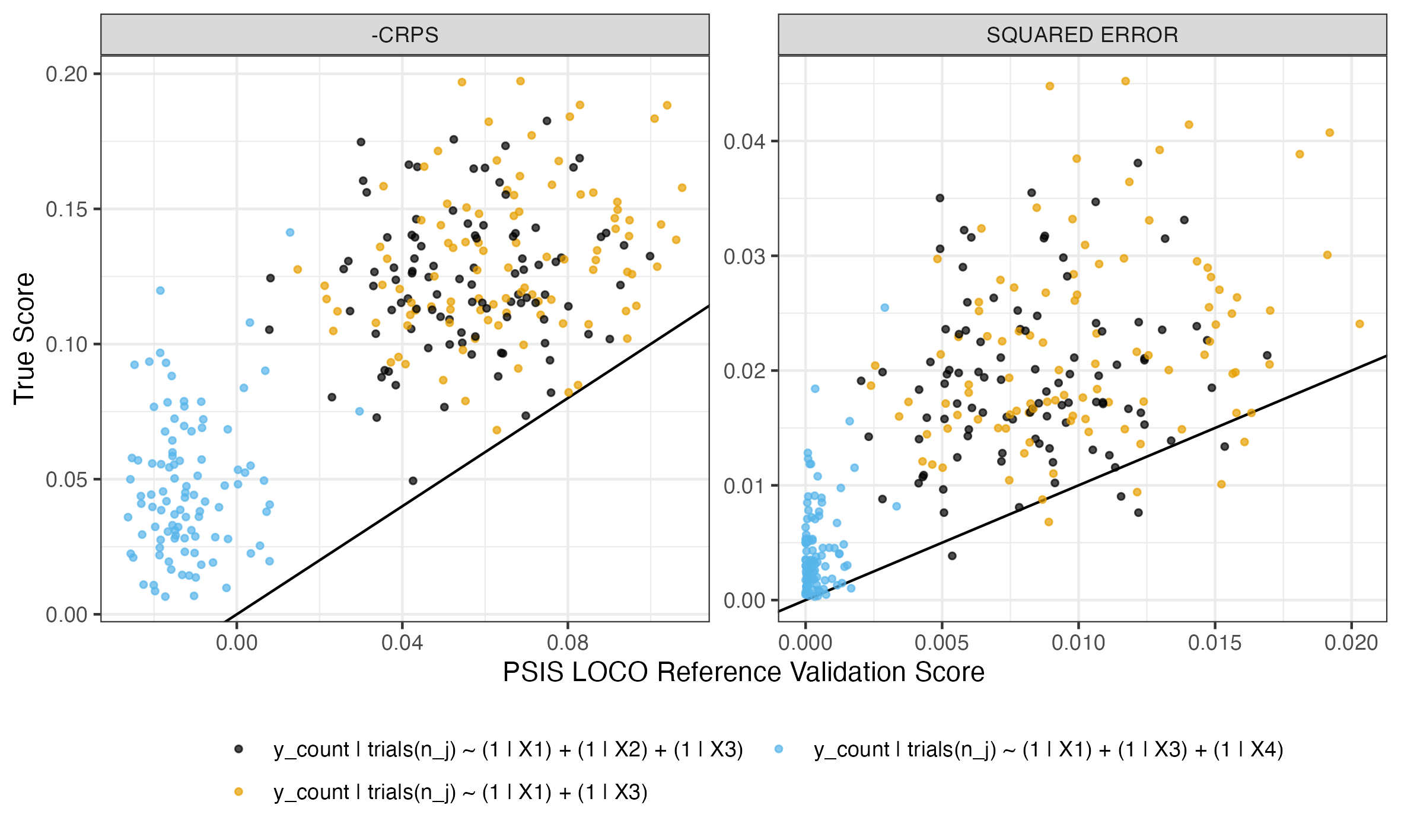}
    \caption{\em Comparison of the true score for the estimate ($y$-axis) against the combined validation approach ($x$-axis). Colour of point represents the different candidate models, whilst different points within this represent different simulation iterations. Points below the black line represent underestimation of the magnitude for CRPS and above the identity line represent underestimation of the magnitude for squared error. }
    \label{fig:combined_ref_model}
\end{figure}

In Figure~\ref{fig:combined_ref_model} we demonstrate the efficacy of this approach against the true model score. Similar to previous results, the magnitude of the score is underestimated for both scores but ordering is maintained.

\section{Conclusion}
We have consider the challenge of assessing model adequacy and selection in the case where models are being used to predict and then aggregate. Our specific focus is on multilevel regression and poststratification (MRP), but this challenge is relatively common across a range of fields. We propose adaptions to the squared error and continuous ranked probability score (CRPS) and demonstrate that this adaption correctly recovers model ordering at the population and subpopulations, suggesting a successful tool for selecting models. We also provide an approximate leave-one-cell-out approach for fast estimation. We propose a reference validation approach and a combined validation approach for use when not all cells are observed in the sample. Together we feel this provides encouraging progress towards MRP model validation and consider this a considerable improvement on previous model validation attempts (which do not retain correct ordering). 

This work provides evidence that failure to correctly recover model ordering is based in the sum of individual scores. However, our method does consistently underestimate the magnitude of the loss, which should be further investigated. Until a solution has been identified, the magnitude of loss should not be directly used as an approximation of true loss, suggesting that caution should be used when evaluating model adequacy. This work used the simulation design proposed by \cite{kuh2022using}. Future work could consider different sampling conditions and model comparisons (including comparisons of different priors) to ensure generalisability as this scenario is severe and may not be illustrative of real world applications.

\bibliographystyle{agsm}

\bibliography{bib}

\end{document}


\section*{Supplementary}

\section{Demonstration of equivalence}

Here we demonstrate the equivalency of the cellwise decomposition to the original population score. To achieve this, we directly calculate the squared error and CRPS for the MRP estimate at the population level and then also calculate this using the cellwise estimates. 

\begin{figure}[b!]
    \centering
    \includegraphics[width=.9\textwidth]{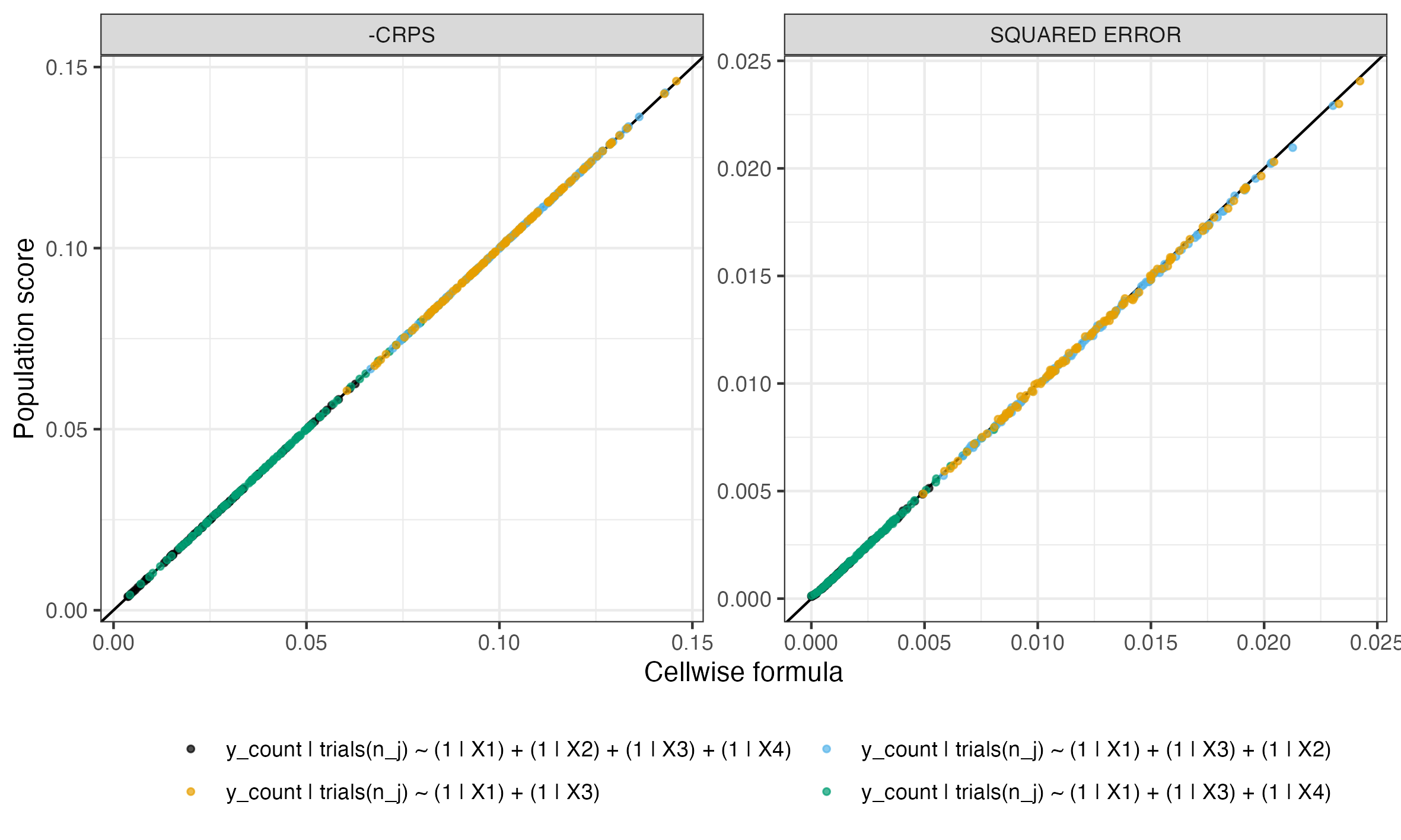}
    \caption{\em Comparison of a direct score of the MRP estimate (using known population truth, $y$-axis) against an indirect score of the MRP estimate (using the known population truth for each cell, $x$-axis). Colour of point represents the different models fitted, whilst different points within this represent different simulation iterations. The black line represents an identity line. As the points all fall either directly or with a small amount of variation, we find evidence that this decomposition of the scoring rule is valid.}
    \label{fig:proof_of_score}
\end{figure}

Figure~\ref{fig:proof_of_score} plots the score for our population mean estimated with MRP against the score calculated through the cellwise error.  Up to a small amount of noise, these two scores are equivalent. This provides confirmation for our proposal that the score for a population mean estimate can be decomposed into cell estimates. 

\newpage 

\section{Using sample as a proxy for the population}

\begin{figure}[b!]
    \centering
    \includegraphics[width=.9\textwidth]{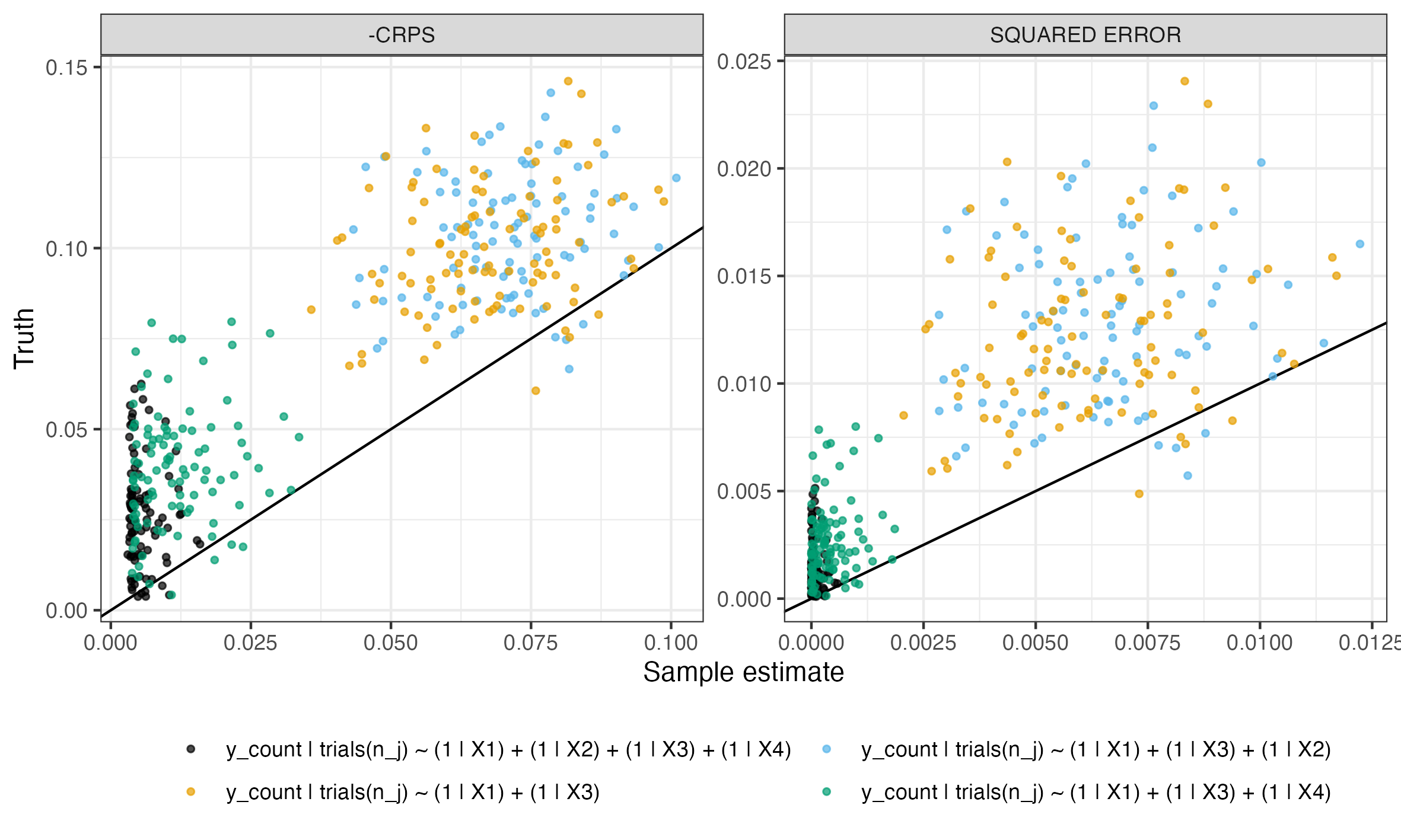}
    \caption{\em Comparison of a direct score of the MRP estimate (using known population truth, $y$-axis) against an indirect score  (using the known sample truth for each cell, $x$-axis). Colour of point represents the different models fitted, whilst different points within this represent different simulation iterations. The black line represents the identity. The CRPS scores estimated with the sample are overall higher than the true score (where lower indicates a worse model) and the squared error scores estimated with the sample are overall lower than the true score (where higher indicates a worse model). This indicates that using the sample as a proxy for population truth underestimates the model error.}
    \label{fig:sample_score}
\end{figure}

Figure~\ref{fig:sample_score} shows comparison of our proposed scores using just the sample (no cross-validation) on the $x$-axis and the true score on the $y$-axis. The error as estimated in the sample is a considerable underestimate relative to the truth. This could be explained by using the sample without correcting for overfitting. 

\newpage
\section{Comparison of scores}

\begin{figure}[b!]
    \centering
    \includegraphics[width=\textwidth]{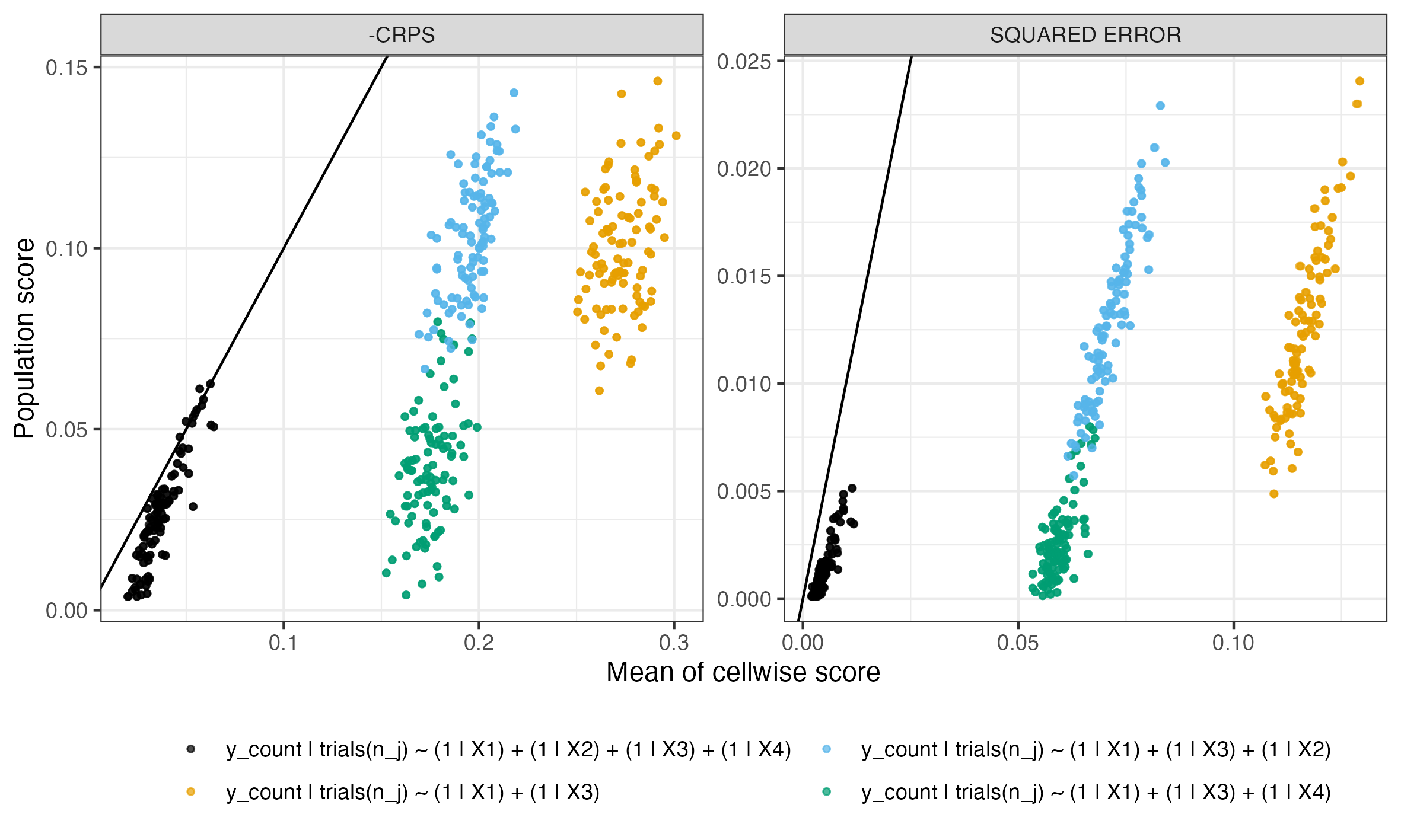}
    \caption{\em Comparison of the true score for the MRP estimate (\textit{y}-axis) the mean cellwise score (\textit{x}-axis). Colour of point represents the different models fitted, whilst different points within this represent different simulation iterations. The black line represents an identity line. Note that mean of scores was take instead of sum of scores (as is more typical) to ensure comparability across simulation iterations. }
    \label{fig:meancellwise_vs_populationtruth}
\end{figure}

Figure~\ref{fig:meancellwise_vs_populationtruth} shows comparison of the traditional mean of cellwise  scores (usually sum of but we use the mean for comparison across simulation iterations) and the MRP population scores. As we can see, the ordering of the models is very different using the mean of cellwise scores when compared to the true ordering. This reflects Kuh et al (2023) but with CRPS and mean squared error scores. 
\newpage
\begin{figure}
    \centering
    \includegraphics[width=\textwidth]{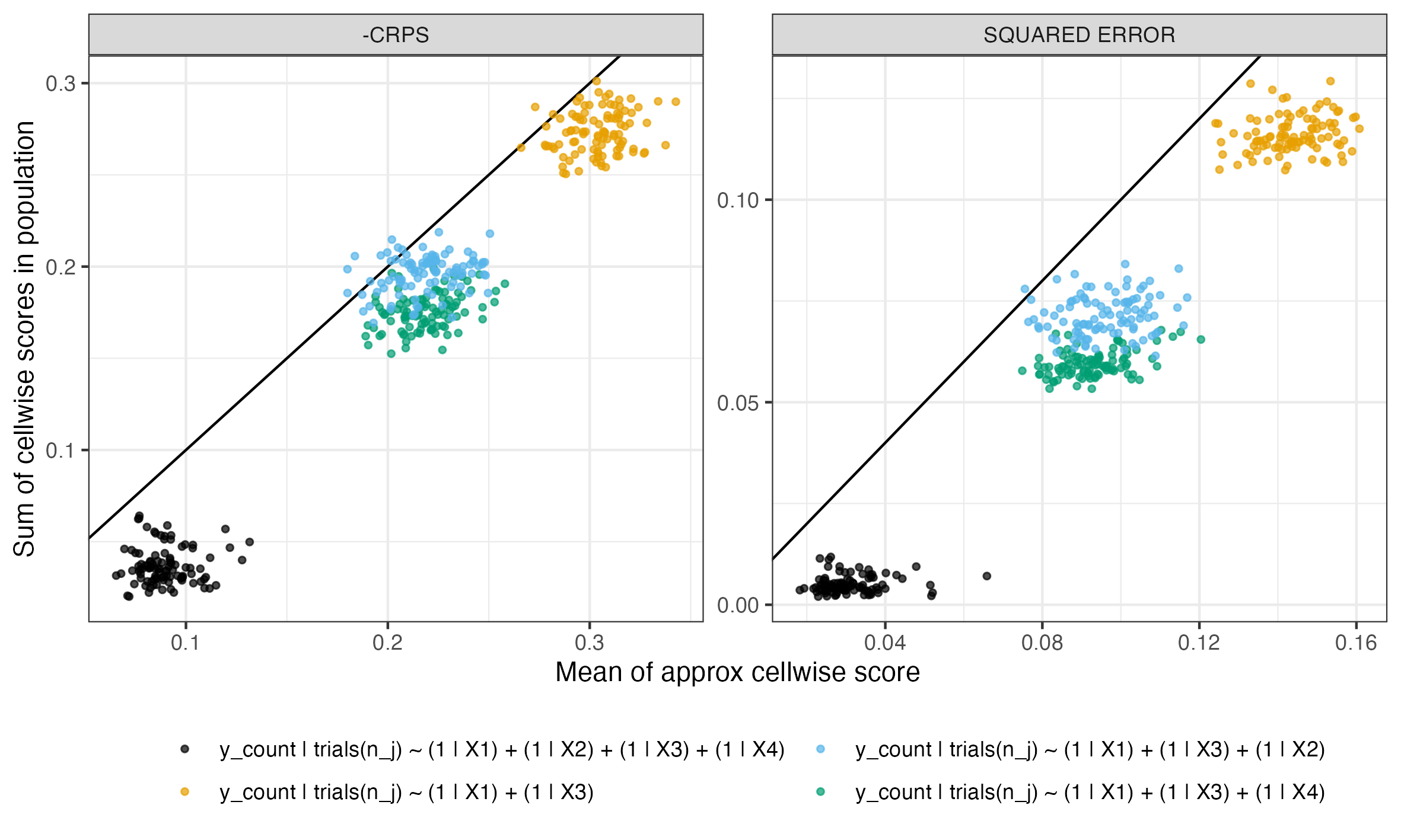}
    \caption{\em Comparison of the the true mean of cell scores in the population (y-axis) against the mean  of PSIS approximated LOCO cell scores (x-axis). Colour of point represents the different models fitted, whilst different points within this represent different simulation iterations. The black line represents an identity line. Note that mean of scores was taken instead of the sum of scores (as is more typical) to ensure comparability across simulation iterations. Interestingly, we see that this approach over-estimates the error (in comparison to our main findings where error was underestimated.}
\end{figure}

\newpage
\begin{figure}
    \centering
    \includegraphics[width=\textwidth]{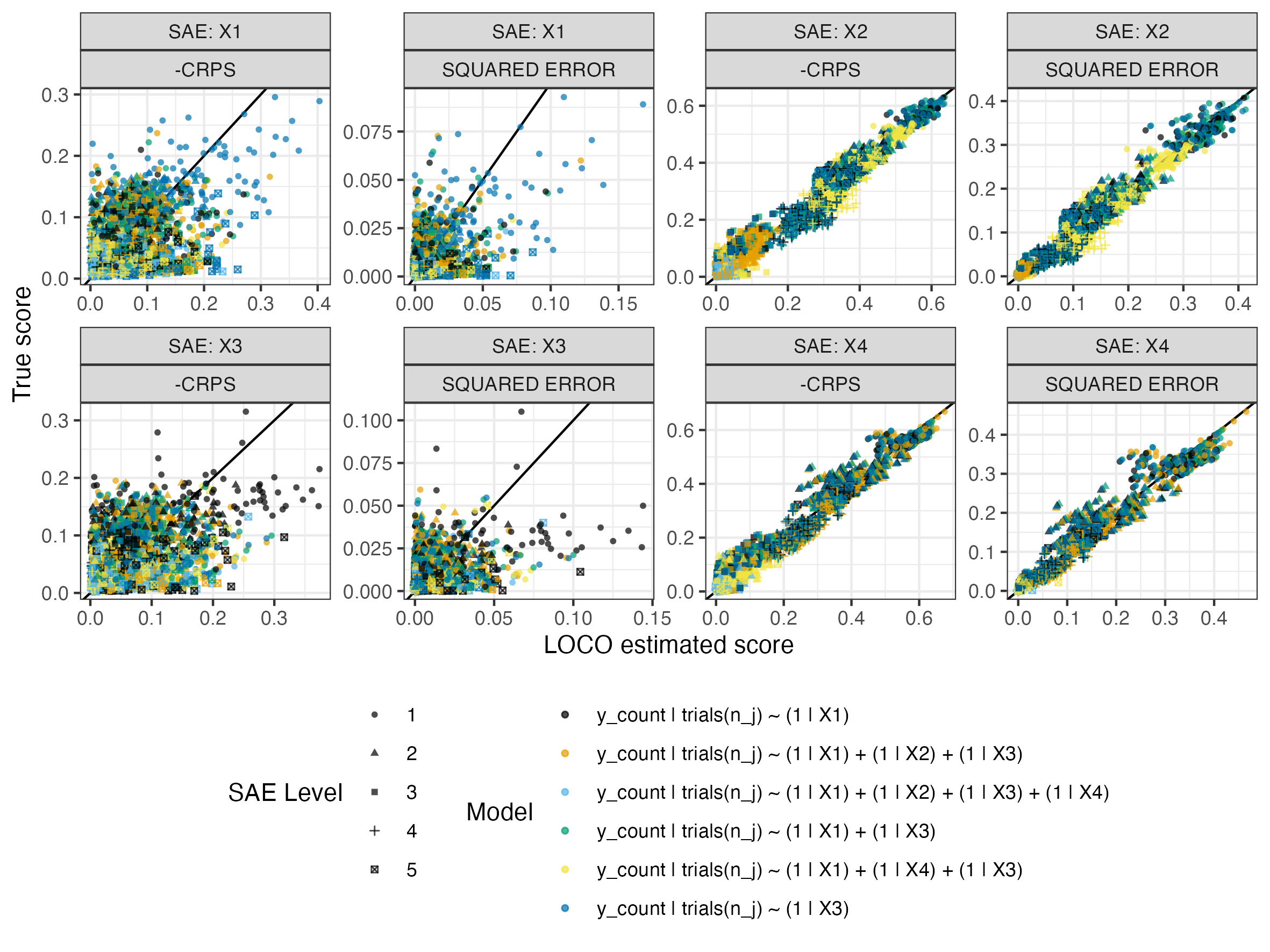}
    \caption{\em Comparison of the true scores across levels (y-axis) against estimated scores (x-axis, vertical facet) for each level in each variable (horizontal facet). Colour of point represents the different models fitted, whilst different points within this represent different simulation iterations. Shape of point represents different levels in the simulation. The black line represents an identity line. }

\end{figure}